\documentclass[12pt,a4paper]{article}

\usepackage{amsmath}
\usepackage{amssymb}
\usepackage{amsthm}
\usepackage[dvipsnames]{xcolor}
\usepackage{dsfont}
\usepackage{esint}
\usepackage{framed}
\usepackage[colorlinks]{hyperref}
\usepackage{lipsum}
\usepackage{mathrsfs}
\usepackage{pgfplots}
\usepackage{slashed}
\usepackage{soul}
\usepackage{upgreek}
\usepackage[all]{xy}
\usepackage[normalem]{ulem}

\newlength{\bibitemsep}
\newlength{\bibparskip}
\let\oldthebibliography\thebibliography
\renewcommand\thebibliography[1]{
	\oldthebibliography{#1}
	\setlength{\parskip}{\bibitemsep}
	\setlength{\itemsep}{\bibparskip}
}

\DeclareMathAlphabet{\mathpzc}{OT1}{pzc}{m}{it}

\newcommand{\rem}[1]{}

\newcommand{\de}{{\rm d}}

\newcommand{\bp}{{\boldsymbol{p}}}

\newcommand{\bX}{{\mathbf{X}}}

\newcommand{\bn}{{\boldsymbol{n}}}
\newcommand{\bzeta}{{\boldsymbol{\zeta}}}

\newcommand{\bsigma}{{\boldsymbol{\sigma}}}

\newcommand{\bz}{{\mathbf{{\boldsymbol{z}}}}}

\newcommand{\bE}{{\mathbf{E}}}
\newcommand{\bF}{{\mathbf{F}}}
\newcommand{\bB}{{\mathbf{B}}}

\newcommand{\be}{\mathbf{e}}

\newcommand{\beq}{\begin{equation}}
\newcommand{\eeq}{\end{equation}}
\newcommand{\ben}{\begin{eqnarray}}
\newcommand{\een}{\end{eqnarray}}

\renewcommand{\contentsname}{}

\numberwithin{equation}{section}

\begin{document}

\title{\vspace{-1 truein}\makebox{Quantum-classical dynamics of Rashba spin-orbit coupling}}
\author{\hspace{-0.7cm}
Paul Bergold{$^1$\footnote{Corresponding author ({\tt paul.bergold@ug.edu.pl})}\,\,,}
Giovanni Manfredi$^2$,
Cesare Tronci$^{3,4}$\smallskip\vspace{-.1cm}
\\
\footnotesize\it\hspace{-1.17cm}\vspace{-.15cm}
$^{1}$Instytut Matematyki, Uniwersytet Gda\'nski, 80--308 Gda\'nsk, Poland
\\
\footnotesize\it\hspace{-1.17cm}\vspace{-.15cm}
$^2$Universit\'e de Strasbourg, CNRS, Institut de Physique et Chimie des Mat\'eriaux de Strasbourg, 67000 Strasbourg, France
\\
\footnotesize\it\hspace{-1.17cm} \vspace{-.15cm}
$^3$School of Mathematics and Physics, University of Surrey, Guildford GU2 7XH, United Kingdom
\\
\footnotesize\it\hspace{-1.17cm} \vspace{-.15cm}
$^4$School of Physical and Mathematical Sciences, Nanyang Technological University, Singapore 637371
\smallskip}

\date{}
\maketitle
\vspace{-1.1cm}
\begin{abstract}
	Mixed quantum-classical models are widely used to reduce the computational cost of fully quantum simulations.
	However, their general applicability across different classes of problems remains an open question.
	Here, we address this issue for systems featuring spin-orbit coupling.
	In particular, we study the interaction dynamics of quantum spin-$1/2$ and classical orbital momentum in one-dimensional models of Rashba nanowires.
	We tackle this problem by resorting to a new quantum-classical Hamiltonian model that, unlike conventional approaches, retains the Heisenberg principle and captures correlation effects beyond the common Ehrenfest method.
	Based on Koopman wavefunctions in classical mechanics, the new model was recently implemented numerically via a particle scheme -- the \emph{koopmon} method -- which is extended here to treat spin-orbit coupling.
	We apply the koopmon method to study the quantum-classical dynamics of nanowire models, with and without the presence of a harmonic potential and in both Rashba-dominated (strong coupling) and Zeeman-dominated (weak coupling) regimes.
	Considering realistic semiconductor parameters, the results are contrasted with both fully quantum and quantum-classical Ehrenfest dynamics.
	In the absence of external potential, the koopmon method qualitatively reproduces the features of the fully quantum evolution for all coupling regimes.
	While it exhibits a slight loss in spin accuracy compared to Ehrenfest simulations, the latter fail to capture the orbital dynamics.
	In the presence of a harmonic potential, the koopmon scheme reproduces several features of full quantum dynamics with accuracy levels that are unachievable by the Ehrenfest model.
	We conclude by presenting a test case that exhibits the formation of cat-like states.
\end{abstract}
\vspace{-1.3cm}
{
\contentsname
\footnotesize
\tableofcontents
}

\section{Introduction}
\subsection{The mixed quantum-classical approach}
The formulation of hybrid models for the dynamics of mixed quantum-classical (MQC) systems has a long history.
In computational chemistry, the challenges posed by fully quantum simulations have stimulated the search for cost-effective MQC codes that treat nuclear motion as classical while retaining the quantum features of the electrons within the molecule \cite{CrBa18}.
While this MQC strategy originally emerged from the adiabatic approximation underlying Born--Oppenheimer molecular dynamics, consistency issues typically arise in treating the full nonadiabatic regime.
For example, popular schemes such as the celebrated \emph{surface hopping} \cite{Bondarenko,Tully} fail to satisfy Heisenberg's uncertainty principle in the quantum sector, while others struggle to properly account for correlation effects, such as quantum decoherence.
Various attempts to overcome these limitations are found in the literature \cite{Aleksandrov,Traschen,Diosi,Gerasimenko,Ghose,Hall,JaSu10,Layton,PeTe}.

In recent years, MQC models have also appeared in other fields including gravity and spintronics.
In the first case, the difficulties in achieving a satisfactory quantum theory of gravity have prompted discussions on whether it is reasonable to accept effective \emph{semiclassical} theories in which classical gravity is coupled to quantum matter \cite{Traschen,Diosi1,Layton}.
Alternatively, in spintronics, atomic spins and magnetic moments are treated classically while the conduction electrons retain fully quantum features \cite{Petrovic}.
In this case, the classical spins are treated via the celebrated Landau--Lifshitz--Gilbert equation \cite{Gilbert} for the average magnetization, which is then coupled to quantum electronic evolution.
The problem of coupling quantum and classical spin systems also arises in chemistry when studying the dynamics of radical pairs \cite{Manolopoulos2}.

MQC models have also appeared under the name of \emph{semiclassical models} \cite{CuEtAl04,HuHeMa17} in the more general setting of solid state physics and spintronics, where orbital degrees of freedom are treated classically while the spin evolution retains its fully quantum features.
In this context, the Eilenberger equation \cite{Eilenberger} typically used in spintronics \cite{Kailasvuori,Halperin,Gorini} is a collisional extension of the quantum-classical Liouville equation (QCLE) \cite{Aleksandrov,Gerasimenko,Kapral} from chemical physics.
While the latter is another widely studied model, it is known to often fail in retaining a positive density matrix at all times, thereby undermining the Heisenberg principle.
A similar failure of positivity also arises in the classical sector \cite{Schofield}.
For example, the QCLE reduces to fully quantum Wigner dynamics for the Rabi model in quantum optics \cite{FoEtAl2019}: while the Heisenberg principle is ensured for this particular case, the corresponding dynamics leads to unsigned phase-space distributions contradicting the classical approximation in the underlying MQC treatment.
Therefore, the feasibility of MQC approaches remains an open question and, in particular, here we aim to overcome consistency issues and provide an understanding of the extent to which MQC models succeed in capturing the correlations arising from spin-orbit coupling.

General configurations couple spin dynamics simultaneously to position and momentum, posing challenges for standard fully quantum methods and motivating the use of MQC approaches.
In the present proof-of-concept study, however, we focus on model validation for the one-dimensional dynamics of semiconductor quantum wires featuring the coupling between spin and orbital momentum only.
Instead of comparing the computational performance of consistent MQC models, here we ask whether the latter can reproduce the spin-orbit correlations in simple situations where the direct comparison with a fully quantum benchmark remains reliable.

\subsection{Ehrenfest quantum-classical model\label{sec:Ehrenfest}}
While several MQC models are available in the literature, here we restrict ourselves to considering two models that succeed in satisfying a series of stringent consistency criteria \cite{Traschen,Terno}, including the Heisenberg uncertainty principle.
As mentioned above, the latter is allowed to fail in general for the standard Boltzmann-type semiclassical models found in the spintronics literature \cite{Eilenberger,Kailasvuori,Halperin,Gorini}.
In addition, the models considered in this paper reproduce \emph{decoherence}, that is, the nontrivial dynamics of quantum purity $\|\hat\rho\|^2=\operatorname{Tr}\hat\rho^2$ representing the transition of the pure quantum state into a mixed state. 

The first model under consideration is the \emph{Ehrenfest model}:
\begin{align}\label{Ehrenfest}
	i\hbar\frac{\partial\widehat{\cal P}}{\partial t}+i\hbar\operatorname{div}\!\big(\widehat{\cal P}\langle\bX_{\widehat{H}}\rangle\big)
	=\big[\widehat{H},\widehat{\cal P}\big].
\end{align}
Here, $\widehat{H}=\widehat{H}(q,p)$ is a Hamiltonian (Hermitian) operator on the quantum Hilbert space $\mathscr{H}$ parameterized by the classical phase-space coordinates.
Also, $\bX_{\widehat{H}}=(\partial_p{\widehat{H}},-\partial_q{\widehat{H}})$ is an operator-valued Hamiltonian vector field, while $\widehat{\cal P}$ is a sufficiently smooth distribution taking values in the space $\operatorname{Her}(\mathscr{H})$ of Hermitian operators on $\mathscr{H}$.
The classical Liouville density and the quantum density matrix are given as
\begin{align*}
	f
	=\operatorname{Tr}\widehat{\cal P}
	\qquad\text{and}\qquad
	\hat\rho
	=\int\!\widehat{\cal P}\,\de q\de p,	
\end{align*}
respectively, so that $\langle\widehat{A}\rangle=\operatorname{Tr}(\widehat{\cal P}\widehat{A})/f$ identifies a local expectation value.
From \eqref{Ehrenfest}, we observe that $\widehat{\cal P}$ undergoes a unitary evolution (due to the commutator) while being Lie-transported (due to the divergence) by the phase-space paths generated by the vector field $\langle\bX_{\widehat{H}}\rangle$.
Consequently, $\widehat{\cal P}$ is also preserved to be Hermitian and positive semidefinite.

In computational chemistry, the model \eqref{Ehrenfest} is used extensively in different variants \cite{Vanicek} due to the computational simplicity of its particle-scheme implementation.
The latter is obtained by first rewriting \eqref{Ehrenfest} as ${\partial_t}f+\operatorname{div}(f\langle\bX_{\widehat{H}}\rangle)=0$ and $i\hbar\partial_t\,\widehat{\!\mathscr{P}}+i\hbar\langle\bX_{\widehat{H}}\rangle\cdot\nabla\,\widehat{\!\mathscr{P}}=[\widehat{H},\,\widehat{\!\mathscr{P}}]$, where $\,\widehat{\!\mathscr{P}}=\widehat{\cal P}/f$ and $\langle\bX_{\widehat{H}}\rangle=\operatorname{Tr}(\,\widehat{\!\mathscr{P}}\bX_{\widehat{H}})$.
Then, the first equation for $f$ is solved by the \emph{point-particle solution ansatz}
\begin{align*}
	f(\bz,t)
	=\sum_{a=1}^Nw_a\delta(\bz-\bzeta_a(t)),
\end{align*}
with $\bz=(q,p)$ and
\begin{align*}
	\bzeta_a(t)
	=\left(q_a(t),p_a(t)\right),\qquad
	\dot{\bzeta}_a
	=\langle\bX_{\widehat{H}}\rangle|_{\bz=\bzeta_a},	
\end{align*}
as well as probability weights $w_a>0$ such that $\sum_aw_a=1$.
Here, the quantity $\langle\bX_{\widehat{H}}\rangle|_{\bz=\bzeta_a}$ requires evaluating ${\hat\varrho_a(t):=\,\widehat{\!\mathscr{P}}(\bzeta_a(t),t)}$ at all times, which can be done by multiplying the $\,\widehat{\!\mathscr{P}}$-equation by $f$ and integrating over phase space, so that $i\hbar\dot{\hat\varrho}_a=[\widehat{H}_a,\hat\varrho_a]$.
The resulting \emph{multi-trajectory Ehrenfest} (MTE) system reads
\begin{align}\label{MFeqs}
	\dot{q}_a
	=\partial_{p_a\!}\langle\hat\varrho_a|\widehat{H}_a\rangle,\qquad\ 
	\dot{p}_a
	=-\partial_{q_a\!}\langle\hat\varrho_a|\widehat{H}_a\rangle,\qquad\ 
	i\hbar\dot{\hat\varrho}_a
	=[\widehat{H}_a,\hat\varrho_a],
\end{align}
with $\widehat{H}_a=\widehat{H}(\bzeta_a)$ and $\langle\hat\varrho_a|\widehat{H}_a\rangle=\operatorname{Tr}(\hat\varrho_a\widehat{H}_a)$.
This scheme is cost-effective from a computational perspective \cite{FaJiSp18} and reproduces decoherence effects.
Indeed, upon using $\hat\rho=\int\!f\,\widehat{\!\mathscr{P}}\de q\de p$, we notice that the quantum purity $\|\hat\rho\|^2=\|\sum_a w_a\hat\varrho_a\|^2$ undergoes nontrivial dynamics.
Despite its wide use in MQC simulations, the MTE system is known to struggle in reproducing accurate results, meaning that the dynamics predicted by \eqref{MFeqs} may deviate substantially from that obtained by the fully quantum description, both quantitatively and qualitatively.
Various ad hoc fixes have been proposed \cite{AkLoPr14} and much of the community has been working to identify alternative MQC approaches retaining a better accuracy \cite{CrBa18}.

\subsection{Koopman model and the \emph{koopmon} scheme\label{sec:koopmons}}
In an attempt to go beyond Ehrenfest dynamics, our team has proposed a new model that is based on the properties of Koopman wavefunctions \cite{Koopman}.
The latter are square-integrable functions $\chi(\bz,t)$ on phase space such that their modulus squared $f=|\chi|^2$ satisfies the classical Liouville equation $\partial_t f=\{H,f\}$.
The general evolution of $\chi$ is given by $i\hbar\partial_t\chi=i\hbar\{H,\chi\}+\varphi\chi$, where the choice of phase $\varphi$ leads to different variants of Koopman wavefunction dynamics.
The main observation is that the operator $i\hbar\{H,\_\,\}$ is self-adjoint for wavefunctions that decay sufficiently fast, so that classical mechanics can be formulated as linear unitary evolution on a Hilbert space, in analogy with standard quantum mechanics.
Instead of taking classical limits, the quantum-classical \emph{Koopman model} presented in this section was obtained by starting with two fully classical systems and then quantizing one of them \cite{BoGBTr19,GBTr20}.
This process leads to a linear \emph{quantum-classical wave equation} \cite{BoGBTr19,Manfredi23} for a hybrid quantum-classical wavefunction $\Upsilon(\bz,x,t)$ in the tensor-product space of Koopman and Schr\"odinger wavefunctions (here, $x$ denotes the spatial coordinate of the quantum subsystem).
Later, a symmetry principle was applied in such a way that the invariance of the dynamics under classical phase transformations $\Upsilon(\bz,x)\mapsto e^{i\varphi(\bz)/\hbar}\Upsilon(\bz,x)$ leads to the Noether quantity $f=\int|\Upsilon|^2\,\de x$, identifying a positive-definite classical density \cite{GBTr21}.
This symmetry principle leads to a nonlinear system, whose full construction was detailed in \cite{GBTr22,GBTr21} and further summarized in \cite{TrGB23}.

Upon restricting to a finite-dimensional quantum Hilbert space $\mathscr{H}=\Bbb{C}^n$ and denoting $\widehat{\cal P}(\bz)=\Upsilon(\bz)\Upsilon(\bz)^\dagger$ so that $f=\operatorname{Tr}\widehat{\cal P}$, the final form of the Koopman model reads
\begin{align}\label{HybEq1}
	i\hbar\frac{\partial\widehat{\cal P}}{\partial t}+i\hbar\operatorname{div}\!\big(\widehat{\cal P}\boldsymbol{\cal X}\big)
	=\big[\widehat{\cal H},\widehat{\cal P}\big],	
\end{align}
with
\begin{align}\label{HybEq2}
	\boldsymbol{\cal X}
	=\langle\bX_{\widehat{H}}\rangle+\frac\hbar{2f}\operatorname{Tr}\!\big(\bX_{{\widehat{H}}}\cdot\nabla\boldsymbol{\widehat{\Sigma}}-\boldsymbol{\widehat{\Sigma}}\cdot\nabla\bX_{{\widehat{H}}}\big),
	\qquad\qquad
	\boldsymbol{\widehat{\Sigma}}
	=\frac{i}f\big[\widehat{\cal P},\bX_{\widehat{\cal P}}\big],	
\end{align}
and
\begin{align}\label{HybEq3}
	\widehat{\mathcal{H}}
	=\widehat{H}+\frac{i\hbar}f\left[\nabla\widehat{\cal P}-\frac1{2f}\widehat{\cal P}\nabla f,\bX_{\widehat{H}}\right].
\end{align}
We observe that both the vector field $\boldsymbol{\cal X}$ in \eqref{HybEq2} and the effective Hamiltonian \eqref{HybEq3} are $\hbar$-corrections of the corresponding quantities appearing in the Ehrenfest model \eqref{Ehrenfest}, that is $\langle\bX_{\widehat{H}}\rangle$ and $\widehat{H}$, respectively.
One may shed some light on the nature of these correction terms by writing \eqref{HybEq1}-\eqref{HybEq3} in their Hamiltonian form, that is \cite{GBTr21,TrGB23}
\begin{align}\label{HamEq}
	i\hbar\frac{\partial\widehat{\cal P}}{\partial t}+i\hbar\operatorname{div}\!\big(\widehat{\cal P}\big\langle\bX_{\delta h/\delta\widehat{\cal P}}\big\rangle\big)
	=\bigg[\frac{\delta h}{\delta\widehat{\cal P}},\widehat{\cal P}\bigg],
	\qquad\qquad
	h(\widehat{\cal P})
	={\int}\big\langle{\widehat{H}}(\operatorname{Tr}\widehat{\cal P})+i\hbar\{\widehat{\cal P},\widehat{H}\}\big\rangle\,\de^2z,	
\end{align}
where the Hamiltonian functional $h$ identifies the total energy of the system and $\delta h/\delta\widehat{\cal P}$ denotes its functional derivative.
The explicit Poisson-bracket structure associated to \eqref{HamEq} was presented in \cite{GBTr21,TrGB23}.
The energy density $\langle{\widehat{H}}(\operatorname{Tr}\widehat{\cal P})+i\hbar\{\widehat{\cal P},\widehat{H}\}\rangle$ comprises a first term $f\langle\widehat{H}\rangle$ and an $\hbar$-correction $\langle i\hbar\{\widehat{\cal P},\widehat{H}\}\rangle$ carrying inhomogeneities that produce correlation effects.
While the first term alone would reduce the $\widehat{\cal P}$-equation in \eqref{HamEq} to the Ehrenfest model \eqref{Ehrenfest}, the extra term in $h(\widehat{\cal P})$ carries the operator $i\hbar\{\widehat{H},\_\,\}$ that is produced by the Koopman approach \cite{BoGBTr19,GBTr22}.
This second term in the integrand is responsible for the \emph{quantum backreaction} on the classical system and we refer to the corresponding integral as the \emph{backreaction energy}.
At present, the specific physical content of the expression of the backreaction energy remains an open question, although we have recently unfolded some analogies with spin-orbit coupling in \cite{Tronci26}.

Recovering Born--Oppenheimer theory in the adiabatic approximation \cite{BeTr23}, the model \eqref{HybEq1}-\eqref{HybEq3} was recently extended to consider MQC spin systems \cite{GBTr23} and solute-solvent interaction in classical fluid solvents \cite{GBTr-fluid,GBTr-fluid2}.
Its Heisenberg picture is available in \cite{DMCTr24} and the quantum-classical entropy functionals for both Ehrenfest and Koopman MQC dynamics were presented in \cite{TrMCGB}.
As discussed in \cite{TrGB23}, despite the formidable look of equations \eqref{HybEq1}-\eqref{HybEq3}, this system does not involve differentials of order higher than two.
Nevertheless, its high level of complexity makes the system hardly tractable by standard methods.
In order to formulate and implement a numerical scheme, in \cite{Bauer24} we presented a particle method that is obtained by exploiting the variational structure underlying the continuum model \eqref{HybEq1}-\eqref{HybEq3}.
In the present paper, we follow the analogous procedure at the level of the Hamiltonian structure in \eqref{HamEq}.
The overall strategy consists of regularizing the backreaction energy density, that is, replacing ${f}^{-1}\operatorname{Tr}(\widehat{\cal P}\{\widehat{\cal P},\widehat{H}\})\longrightarrow\bar{f}^{-1}\operatorname{Tr}(\bar{\cal P}\{\bar{\cal P},\widehat{H}\})$, where $f=\operatorname{Tr}\widehat{\cal P}$.
Here, the bar-symbol denotes convolution with a smooth, strictly positive convolution kernel $K^{(\alpha)}(\bz)$ on phase space, which depends on a positive parameter $\alpha>0$ so that the sequence $(K^{(\alpha)})$ converges to the delta distribution in the limit $\alpha\to 0$.
In addition, we ask that $K^{(\alpha)}(\bz)=K^{(\alpha)}(-\bz)$ and we will drop the superscript for ease of notation. 

Upon writing $\bar{\cal P}(\bz,t)={\int}K(\bz-\bz')\widehat{\cal P}(\bz',t)\,\de q'\de p'$ and using ${\bar{f}=\operatorname{Tr}\bar{\cal P}}$, the regularized Hamiltonian functional $\bar{h}(\widehat{\cal P}):=\operatorname{Tr}{\int}\big(\widehat{\cal P}{\widehat{H}}+i\hbar\bar{\cal P}\{\bar{\cal P},\widehat{H}\}/{\operatorname{Tr}}\bar{\cal P}\big)\,\de^2z$ transforms the $\widehat{\cal P}$-equation into
\begin{align*}
	i\hbar\frac{\partial\widehat{\cal P}}{\partial t}+{i\hbar}\operatorname{div}\big(\widehat{\cal P}\langle\bX_{\delta\bar{h}/\delta\widehat{\cal P}}\rangle\big)
	=\bigg[\frac{\delta\bar{h}}{\delta\widehat{\cal P}},\widehat{\cal P}\bigg].
\end{align*}
Importantly, this regularized functional $\bar{h}(\widehat{\cal P})$ remains finite upon replacing the ansatz $\widehat{\cal P}(\bz,t)=\sum_aw_a\hat\varrho_a(t)\delta(\bz-\bzeta_a(t))$.
Indeed, we observe that the first term of the total energy $h(\widehat{\cal P})$ in \eqref{HamEq} does not require any regularization by itself.
Notice that the singular solution ansatz on $\widehat{\cal P}(\bz,t)$ differs from the usual mean-field product-state factorization between the classical and quantum sectors.
Indeed, quantum-classical correlations are still retained by the dependence of the density matrix $\hat\varrho_a(t)$ on the particle label $a$ and by the backreaction force terms.
At this point, the same procedure leading to the MTE equations \eqref{MFeqs} can be applied to give the \emph{koopmon} scheme \cite{Bauer24,TrGB23}:
\begin{align}\label{MFeqs2}
	\dot{q}_a
	=w_a^{-1}{\partial_{p_a\!}h},\qquad
	\dot{p}_a
	=-w_a^{-1}{\partial_{q_a\!}h},\qquad
	i\hbar{\dot{\hat\varrho}_a}
	=w_a^{-1}[{\partial_{\hat\varrho_a\!}h},\hat\varrho_a],	
\end{align}
where
\begin{align}\label{koopmonHam}
	h
	=\sum_aw_a\langle\hat\varrho_a,{\widehat{H}}_a\rangle+\frac12\sum_{a,b}w_aw_b\big\langle i\hbar[{\hat\varrho}_a,{\hat\varrho}_b],\widehat{\cal I}_{ab}\big\rangle,	
\end{align}
and we denote
\begin{align}\label{koopint}
	\widehat{\cal I}_{ab}
	:=\frac12\int\frac{K_a\{K_b,\widehat{H}\}-K_b\{K_a,\widehat{H}\}}{\sum_c w_c K_c}\,\de^2z,
	\qquad\text{with}\qquad
	K_s(\bz,t)
	:=K(\bz-\bzeta_s(t)).	
\end{align}
Just as the hybrid PDE \eqref{HybEq1}-\eqref{HybEq3} can be viewed as an $\hbar$-correction to the Ehrenfest PDE \eqref{Ehrenfest}, the above koopmon scheme can be viewed as an $\hbar$-correction to the MTE method \eqref{MFeqs}.
As explained in \cite{Bauer24}, this particle closure comprises a Hamiltonian system and, as a result, retains fundamental conservation laws such as energy and total probability.
This structure also preserves the nonnegative classical weights, the Hermiticity and positivity of the particle density matrices, provided these properties hold initially.
Thus, even when the approximation loses quantitative accuracy in strongly non-classical regimes, the particle representation continues to ensure non-negative classical and quantum probabilities.
An analogous structure-preserving particle scheme was also proposed in the context of quantum hydrodynamics \cite{FoHoTr19,HoRaTr21}.

The computational cost of the koopmon scheme depends mainly on the value of the regularization parameter $\alpha$.
Small values of $\alpha$ require a fine phase-space grid for the evaluation of the backreaction terms, while larger values reduce the cost.
In particular, in the limit $\alpha\to\infty$ the kernel $K^{(\alpha)}$ tends to zero and its derivatives vanish.
All backreaction terms then drop out, and the koopmon method converges to the MTE method \eqref{MFeqs}.
Previous numerical experiments show that values $\alpha\gtrapprox 10$ produce close agreement between the koopmons and Ehrenfest dynamics.
We emphasize, however, that this agreement is structural rather than a rigorous asymptotic error estimate.

The koopmon scheme \eqref{MFeqs2}-\eqref{koopint} was successfully validated in \cite{Bauer24} for typical benchmark problems in physics and molecular dynamics.
In the first case, we considered the Rabi problem for the interaction of a quantum $1/2$-spin with a classical harmonic oscillator in both ultrastrong and deep-strong coupling regimes.
In the second case, we considered typical nonadiabatic problems that are well-known in computational chemistry (\emph{Tully models}) \cite{Tully90}.
In all these cases, the quantum-classical coupling is realized by an interaction potential depending only on the classical position coordinate.
Conversely, typical problems in spintronics involve spin-momentum interaction and, more generally, the simultaneous interaction of quantum spins and both position and momentum in the form of the spin-orbit coupling operator discussed below.

This simultaneous coupling between position and momentum challenges numerical operator-splitting approaches to the time-dependent Schr\"odinger equation~\cite{Lubich08}, since standard Fourier-based techniques are not applicable.
In this case, one typically resorts to computationally expensive approaches, such as Krylov subspace methods (e.g. Lanczos iterations) \mbox{\cite{PaLi86,HoLu97,Lubich08,IsKrSi19}}.
To avoid this level of difficulty in the generation of a fully quantum benchmark, here we assess MQC models for systems in which the spin is coupled only to the orbital momentum, as in the case of quantum nanowires \cite{MiKi01,GoZu02,GoZu04}.
This allows us to resolve the quantum dynamics by standard Fourier-based methods for wavepacket propagation \cite{Hardin73,Feit82,Kosloff83}.
In particular, we present a proof-of-concept validation study of MQC Ehrenfest and Koopman dynamics for spin-momentum correlations in the presence of Rashba spin-orbit coupling in one dimension.
The following section offers a general introduction to Rashba dynamics and its one-dimensional approximations.

\subsection{Spin-orbit coupling and Rashba nanowire models\label{sub:Spin-orbit coupling and Rashba dynamics in nanowire models}}
The spin-orbit coupling (SOC) is a relativistic effect that arises when an electron moves in an electric field -- typically the electric field generated by the nucleus in an atom.
In the reference frame of the electron, the electric field also possesses a magnetic component, which interacts with the spin of the electron.
The semi-relativistic SOC Hamiltonian can be derived from the Dirac equation, by making use of the Foldy--Wouthuysen transformation \cite{Foldy,Thomas} to expand the Dirac equation in powers of the inverse of the speed of light $c$.
To second order in $1/c$, the SOC Hamiltonian operator reads as $\widehat{H}_\text{SOC}=\hbar\widehat\bsigma\times\bF\cdot\hat{\bp}/(2m_ec)^2$ \cite{Hervieux}.
Here, $\hat{\bp}=-i\hbar\nabla$ is the quantum canonical momentum and $\bF=-e\bE$ is the force corresponding to the electric field $\bE$, $m_e$ is the electron mass, and $e>0$ is the absolute value of the electron charge.
In the present work, we consider a simplified SOC term given by the Rashba Hamiltonian $\widehat{H}_\text{R}=\alpha_\text{R}\,\widehat\bsigma\times\be_z\cdot\hat{\bp}$ \cite{ByRa84}, which can be obtained formally from the above SOC Hamiltonian by taking a uniform electric field in the direction $\be_z$.
Note that the Rashba coupling parameter $\alpha_\text{R}$ has the dimensions of a velocity.
The Rashba Hamiltonian can be rewritten as $\widehat{H}_\text{R}=\mu_B\bB_\text{R}\cdot\widehat\bsigma$, with $\mu_B$ the Bohr magneton, and $\bB_\text{R}=\alpha_\text{R}(\be_z\times\hat{\bp})/\mu_B$ may be interpreted as an effective momentum-dependent magnetic field.
In a semiconductor quantum well, structural inversion symmetry is broken along the growth direction, producing an interfacial electric field perpendicular to the well, which leads to the Rashba Hamiltonian described above \cite{Manchon15}.

\begin{figure}
	\begin{framed}\noindent
		\centering
		\includegraphics[width=10.5cm]{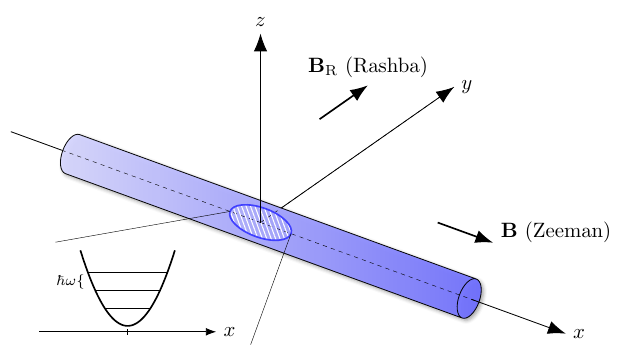}
		\vspace{-.25cm}
	\end{framed}
	\noindent\vspace{-.9cm}
	\caption{\footnotesize
		Sketch of a quantum nanowire with its axis along the $\be_x$ direction.
		The Rashba magnetic field points along the $\be_y$ axis, while the external magnetic field $\bB$ (Zeeman effect) is directed along the axis of the nanowire.
		In the case of a non-ballistic nanowire, a quantum dot is modeled by a harmonic potential with energy level spacing $\hbar\omega$.
	}\label{0-model}
\end{figure}

As we are interested in semiconductor quantum nanowires, we want to derive the Rashba Hamiltonian for a one-dimensional (1D) system, which is usually done as follows \cite{BaBe24,NPFrBaKo10,GoZu02,MiKi01}.
Starting from the Rashba Hamiltonian $\widehat{H}_\text{R}$ for a 2D electron gas confined to the $(x,y)$ plane, a strong transverse confinement quantizes the motion in the direction $\be_y$ perpendicular to the wire axis $\be_x$, and produces a discrete set of subbands.
When the confinement energy, separating the lowest transverse mode from higher modes, exceeds all other relevant energy scales -- chemical potential, Zeeman energy, temperature, etc. -- the system effectively occupies only the lowest subband.
Projecting the full 2D Hamiltonian onto this mode eliminates the transverse momentum operator, whose expectation value vanishes by symmetry, leaving only the longitudinal momentum as a dynamical degree of freedom \cite{GoZu02,MiKi01}.
This procedure holds when the wire width is much smaller than the Rashba length $L=m_*^{-1}\hbar/\alpha_\text{R}$, and yields the 1D Rashba Hamiltonian $\widehat{H}_\text{R}=\alpha_\text{R}\,\widehat{\sigma}_{y}\hat{p}_x$.
Note that $m_*$ denotes the effective electron mass.
Other examples of the application of the 1D Rashba Hamiltonian to quantum nanowires can be found in Refs. \cite{BiMe08,GoRoDo22,PeMi05,StSe03}.
One can further include an external magnetic field $\bB=B_x\be_x$ oriented along the nanowire, as done, for instance, in \cite{DoRo18}, as well as an external harmonic potential with trapping frequency $\omega$, mimicking a quantum dot \cite{Fan16} (see Figure~\ref{0-model}).
The resulting 1D Hamiltonian is given by
\begin{align}\label{BallHam}
	\widehat{H}(\hat q,\hat p)
	=\frac{\hat{p}^2}{2m}+\alpha_\text{R}\widehat{\sigma}_y\hat p+\frac{1}{4}g_eB_x\widehat{\sigma}_x+\frac{1}{2}m\omega^2{\hat q}^2,	
\end{align}
where we relabel $x\to q$ and $\hat p_x\to \hat p$ the longitudinal position and momentum variables along the nanowire, respectively.
Here, quantities are expressed in atomic units, for which $\hbar=1$ (reduced Planck constant), $m_e=1$ (electron mass), $e=1$ (absolute electron charge), and $\mu_B={1}/{2}$ (Bohr magneton).
Hence, $m$ represents the effective electron mass $m_{*}$ in terms of the real electron mass ($m=m_{*}/m_e$), $\hat p=-i\partial_q$ denotes the momentum operator along the wire, and $g_e$ is the Land\'e $g$-factor of the material.
In these units, the Zeeman energy is written as
\begin{align*}
	E_{\text{Z}}
	:=\frac{1}{2}\mu_\text{B}g_eB_x 
	=\frac14 g_e B_x,
\end{align*}
while the harmonic oscillator energy is simply $E_{\rm HO}=\omega$.
When the harmonic oscillator term is absent, the nanowire is called \emph{ballistic}.
Conversely, throughout this paper the term \emph{non-ballistic} refers only to longitudinal harmonic confinement modeling a quantum dot.

In the remainder of this work, we will use the Hamiltonian \eqref{BallHam} as a testbed for our hybrid koopmon approach, and compare the results to fully quantum simulations of the Schr\"odinger equation, as well as simulations based on the Ehrenfest model.
We will not consider cases where the Rashba parameter $\alpha_\text{R}$ is space-dependent -- which may result from the doping profile in a semiconductor \cite{DoRo18} -- as this would introduce additional complications originating from the non-commutativity of the quantum operators $\alpha_\text{R}(\hat q)$ and $\hat p$.
Finally, we mention that a Hamiltonian operator similar to \eqref{BallHam} also appears in the dynamics of spinor Bose--Einstein condensates with SOC, although in that case \eqref{BallHam} is completed by the addition of nonlinear terms as customary in that context \cite{LiJiSp11,ZhMoBuEnZh16}.

\section{Numerical implementation}
In this section, we extend the present version of the koopmon method to include momentum coupling.
The first implementation, originally introduced in \cite{Bauer24}, was limited to hybrid Hamiltonians in which the coupling depends only on the classical position.
We have therefore extended our code so that it applies to general Hamiltonians with coupling in both position and momentum.
The overall structure of the algorithm remains unchanged, but additional backreaction terms must be evaluated.
To give a clear overview of the changes, we begin by recalling the main components of the method and the parameters that influence its accuracy.

We briefly recall the main ingredients of the koopmon method.
The method is controlled by two main parameters:
(i) the number $N>1$ of koopmons, and
(ii) the regularization width $\alpha>0$, which enters the scheme through the regularization kernel $K^{(\alpha)}$ as described in the previous section.
Following \cite{Bauer24}, in this paper we use the Gaussian kernel
\begin{align}\label{kernel}
	K^{(\alpha)}(q,p)
	=\tilde K(q)\tilde K(p)
	\qquad\text{with}\qquad
	\tilde K(y)
	:=\frac{1}{\alpha\sqrt{\pi}}\,e^{-y^2/\alpha^2},
\end{align}
which converges (in the usual sense of distributions) to the delta distribution as $\alpha\to 0$.
Recall that in this limit, the koopmon equations \eqref{MFeqs2} recover the original PDE model \eqref{HybEq1}-\eqref{HybEq3}.

Earlier numerical experiments in \cite{Bauer24} indicate that $N=500$ and $\alpha=0.5$ form a robust pair of parameters, which will be adopted throughout this work.
While the accuracy of the koopmon method can be improved by decreasing $\alpha$ and simultaneously increasing the number of koopmons $N$, this strategy leads to a higher computational cost.
Such a systematic optimization of $(N,\alpha)$ and its associated convergence study lie beyond the purpose of the present work.
Instead, we use the parameter pair previously tested in \cite{Bauer24} for Rabi dynamics to specifically focus on the behavior of MQC dynamics for momentum-dependent Rashba coupling.

\subsection{Koopmon scheme for momentum coupling}
The original implementation of the koopmon method contains two main subroutines.
The first subroutine is the numerical evaluation of the backreaction integral $\widehat{\cal I}_{ab}$ appearing in the equations of motion \eqref{MFeqs2}.
This part of the code is unchanged from \cite{Bauer24}.
More precisely, we continue to use the composite trapezoidal rule on a phase-space box that adapts dynamically to the motion of the koopmons.
Details of the quadrature parameters can be found in Section~3.1 of \cite{Bauer24}.
The second subroutine is the time integration of the full Hamiltonian system \eqref{MFeqs2}.
This part of the implementation is also unchanged.
All simulations presented in this paper use the classical fourth-order Runge--Kutta (RK4) method with a fixed step size $dt=T/400$, where $T>0$ denotes the final simulation time.
We observed that this choice of time step provides sufficient temporal resolution for the simulations considered here.
We also note that a fully time-adaptive integrator and a fourth-order symplectic two-stage RK method have recently been implemented within our group.
These integrators may help reduce computational cost in future work.
In the present context, the adequacy of the fixed RK4 step is assessed through conservation diagnostics for the koopmon Hamiltonian.
In particular, for the test cases considered here, the maximum relative drift $\max_{0\leq t\leq T}|h(t)-h(0)|/|h(0)|$ of the koopmon energy \eqref{koopmonHam} remained below $3.8\times10^{-3}$ over all five test cases, thereby confirming reliability of the RK4 time discretization.

The key novelty of the updated implementation lies in extending the code to enable simulations of Hamiltonians with momentum coupling.
Since the kernel function is chosen in the product form specified in \eqref{kernel}, the first Poisson bracket appearing in the numerator of the backreaction integral $\widehat{\mathcal I}_{ab}$ in \eqref{koopint} can be written as
\begin{align*}
	\{K_b(\bz),\widehat H(\bz)\}
	&=\partial_q K_b(\bz)\partial_p\widehat H(\bz)-\partial_p K_b(\bz)\partial_q\widehat H(\bz)\\
	&=\tilde K_{b}'(q)\tilde K_{b}(p)\partial_p\widehat H(\bz)-\tilde K_{b}(q)\tilde K_{b}'(p)\partial_q\widehat H(\bz).
\end{align*}
An analogous expression holds for the second Poisson bracket $\{K_a,\widehat{H}\}$, also giving rise to two contributions, one involving $\partial_p\widehat{H}$ and the other $\partial_q\widehat{H}$.
In the original implementation, momentum coupling was absent, and only the latter contribution needed to be implemented.
Indeed, this is due to the fact that the backreaction energy $h$ in \eqref{koopmonHam} vanishes identically whenever $\widehat{H}$ is either a purely classical phase-space function or a purely quantum operator.
The updated code now includes the previously omitted backreaction contributions associated with $\partial_p\widehat{H}$, while omitting terms that vanish identically.
More precisely, for pure momentum coupling, all position-coupling terms are suppressed; for pure position coupling, all momentum-coupling terms are suppressed; for Hamiltonians featuring both types of coupling, all backreaction integrals are evaluated.

In later sections, we will compare both the koopmon scheme and MTE to the fully quantum dynamics of Rashba nanowires to assess their relative performance.
Despite the success of MTE in accurately capturing the spin dynamics in some cases, this method struggles in reproducing several essential features that are instead captured by the koopmon method.

\subsection{Quantum solver}
The solutions of the time-dependent Schr\"odinger equation are computed using the \emph{Split-Operator Fourier Transform} (SOFT) method \cite{Lubich08,Greene17}.
The underlying quantum Hamiltonian is obtained from $\widehat H(q,p)$ by replacing the classical variables with the corresponding quantum operators $\hat q=q$ and $\hat p=-i\partial_q$.
Moreover, the initial wavepacket is chosen as $\Psi_0(q)=\psi_0(q)\,(0,1)\in L^2(\mathbb{R},\mathbb{C}^2)$, where $\psi_0$ is a normalized Gaussian centered at $(\mu_q,\mu_p)\in\mathbb{R}^2$, that is,
\begin{align}\label{eq:initial_GW}
	\psi_0(q)
	=(\gamma/\pi)^{1/4}\exp\left(i\mu_p(q-\mu_q)-{\gamma}(q-\mu_q)^2/2\right),
\end{align}
with $\gamma=(2\sigma_q^2)^{-1}$ for a given $\sigma_q>0$.
As described in \cite{Bauer24}, this initialization matches the sampling used in the koopmon method: for each trajectory, the classical position is sampled from a normal distribution with mean $\mu_q$ and variance $\sigma_q^2$.
Analogously, the classical momentum is sampled from a normal distribution with mean $\mu_p$ and variance $\sigma_p^2$ with $\sigma_p = 1/(2\sigma_q)$.
We note that the above choice of the initial wavefunction $\Psi_0$ yields an initial density matrix for each koopmon that is aligned with $\widehat\sigma_z$, that is, $\hat\rho_0=(0,1)(0,1)^T$.

The SOFT simulations presented in this paper use the time step $dt=T/1000$.
Moreover, the spatial grid contains $2^{12}=4,096$ points, and for each test case we choose the spatial domain $\Omega:=[-q_{\max},q_{\max}]$ so that the $L^2$ norm of the wavefunction is conserved up to an error smaller than $10^{-10}$.
From the time-evolved wavepacket $\Psi(t)$ we then compute the Wigner distribution of the orbital degrees of freedom:
for $\Psi=(\psi_1,\psi_2)^T$, we set $W_\Psi=W_{\psi_1}+W_{\psi_2}$, where
\begin{align*}
	W_{\psi_j}(q,p)
	:=\frac{1}{\pi}\int_{-\infty}^\infty\psi_j^*(q+y)\psi_j(q-y)e^{2ipy}\,\mathrm{d}y,
	\quad
	(q,p)\in\mathbb{R}^2,
\end{align*}
are the diagonal terms of the general Wigner matrix.
In this paper, we use the term \emph{wavepacket} for both the position-space wavefunction and its Wigner distribution.
Notice that, unlike the classical Liouville density, the quantum Wigner distribution can take negative values which indicate purely quantum correlations, typically related to interference effects.
By contrast, the MQC particle methods represent the classical sector by nonnegative particle densities.
To compare the quantum simulations with the two MQC particle methods, we plot the particle clouds together with their smoothed density
\begin{align*}
	D(\bz,t)
	:=(K^{(\Delta)}*f)(\bz,t)
	=\sum_aw_aK^{(\Delta)}(\bz-\bzeta_a(t)),
\end{align*}
where $K^{(\Delta)}$ is a Gaussian of width $\Delta>0$.
This width is a mere visualization parameter and therefore independent of $\alpha$.
We fix $\Delta=0.2$ in all test cases.

In the quantum sector we will plot the evolution of purity $\|\hat\rho(t)\|^2=\operatorname{Tr}\hat\rho(t)^2$, which measures decoherence.
In SOFT simulations, $\hat\rho(t)=\int\Psi(q,t)\Psi(q,t)^\dagger\,\de q$, while for the particle methods we have $\hat\rho(t)=\sum_a w_a\hat\varrho_a(t)$.\\

The total discrepancy between the koopmon simulations and the fully quantum reference may reflect contributions from the MQC model itself, the regularized particle closure, \mbox{finite-$N$} sampling, and time discretization.
We emphasize, however, that identifying these potential error sources does not constitute a rigorous error estimate for the numerical experiments below.
The diagnostics reported here assess selected numerical contributions, while establishing a general approximation theorem is left for future work.

\subsection{Numerical rescaling}
When working in a fixed unit system such as atomic units, the Heisenberg uncertainty principle can lead to an initial Wigner distribution that may be strongly anisotropic.
Indeed, it may be highly elongated in position space and correspondingly very narrow in momentum space, or vice versa.
Such extreme aspect ratios pose numerical challenges, as they require the simultaneous handling of very large numerical values (e.g. positions) and very small ones (e.g. momenta) within the same computation.
To improve numerical stability and to avoid the mixing of vastly different numerical scales, we introduce a suitable rescaling of the physical variables.

Since the fully quantum solver SOFT is based on wavepacket propagation in position space, it suffices to rescale the position variable only.
More precisely, for a given scaling parameter $s>0$, we introduce a rescaled initial wavefunction $\psi_0^{s}(q'):=\sqrt{s}\,\psi_0(sq')$ on the rescaled spatial domain $\Omega'=\Omega/s=[-q_{\max}/s,q_{\max}/s]$, where $\Omega$ was introduced above.
By construction, $\psi_0^{s}$ remains normalized, and a short calculation shows that the corresponding Wigner distribution satisfies $W_{\psi_0^{s}}(q,p)=W_{\psi_0}(sq,p/s)$, i.e., the rescaling induces a linear transformation of the phase-space variables.
In our setting, the Wigner distribution $W_{\psi_0}$ of the initial Gaussian wavepacket defined in \eqref{eq:initial_GW} is given by the product of Gaussian distributions in position and momentum, whose variances scale inversely.
This allows the scaling parameter to be chosen such that the rescaled Wigner distribution becomes isotropic.
In particular, requiring equal position and momentum variances uniquely determines $s=\gamma^{-1/2}=\sqrt{2}\,\sigma_q$.
Based on this rescaling of the initial wavepacket, SOFT is applied on the rescaled domain $\Omega'$.
Note that the discretized momentum variables and the time step remain unchanged in the implementation.

For the particle-based MQC methods, a simultaneous rescaling of both the position and momentum variables is required.
More precisely, using the same scaling parameter as above, we introduce the rescaled classical variables $q'=q/s$ and $p'=s\,p$.
In particular, this rescaling is consistent with the one used in the SOFT implementation, in which the spatial argument of the wavefunction is transformed according to $q=s\,q'$.
In the implementation, the particle coordinates are stored in the rescaled variables, while the Hamiltonian functions entering \eqref{MFeqs2} are evaluated at the corresponding physical variables $q=s\,q'$ and $p=p'/s$.
As in the SOFT implementation, the discretized time step remains unchanged.
Notice that, since the map $(q,p)\mapsto(q/s,sp)$ is invertible and canonical, the rescaling preserves the Hamiltonian phase-space structure.
This rescaling allows the particle variables to be stored and compared on balanced numerical scales.

\section{Numerical test cases on ballistic nanowires}
In this section, we present the numerical results obtained with the new version of the koopmon method and compare them with both the MTE method \eqref{MFeqs} and fully quantum solutions.
All simulations, including plotting routines, are implemented in MATLAB (version 2023b).

Our test cases are organized into two main groups.
Test cases 1 and 2 consider MQC Hamiltonians of the form
\begin{align}\label{eq:ballistic}
	\widehat{H}_{\operatorname{bal}}(p)
	=\frac{1}{2m}p^2+\alpha_\text{R}\widehat\sigma_yp+\frac{1}{4}g_eB_x\widehat\sigma_x.
\end{align}
In spintronics, this type of Hamiltonian is commonly referred to as \emph{ballistic}.
This reflects in the `bal' subscript which was inserted for later convenience.
Note that \eqref{eq:ballistic} is obtained from the fully quantum Hamiltonian in \eqref{BallHam} by replacing the quantum canonical observables with the corresponding classical phase-space coordinates.
In Section~\ref{sec:Numerical test cases on non-ballistic nanowires}, we extend the ballistic Hamiltonian introduced in \eqref{eq:ballistic} by incorporating an external harmonic oscillator.

In each case, we will use realistic material properties as found in common semiconductors to probe MQC Rashba dynamics in different coupling regimes.

\paragraph{Classification of coupling regimes.}
It is useful to classify coupling regimes by comparing the spin-orbit (SO) and Zeeman (Z) energy scales.
In atomic units, we define
\begin{align*}
	E_{\text{SO}}
	:=\frac{1}{2}m\alpha_\text{R}^2
	\quad\text{and}\quad
	E_{\text{Z}}			
	:=\frac{1}{4}g_eB_x.
\end{align*}
Following \cite{GoRoDo22}, we then introduce the dimensionless ratio
\begin{align}\label{eq:defR}
	R
	:=\frac{2E_{\text{SO}}}{|E_{\text{Z}}|},
\end{align}
and classify the regimes as \emph{Zeeman-dominated} ($R< 1$) and \emph{Rashba-dominated} ($R>1$).

In this work, studies designed to highlight Rashba effects (test cases 2 and 4) use \emph{indium arsenide} (InAs), which has the strongest Rashba coupling.
Instead, studies focused on strong magnetic-field effects use materials with smaller $\alpha_\text{R}$, namely \emph{indium antimonide} (InSb) (test cases 1 and 3).
See Table~\ref{table:material} for the specific parameter values for each material, along with the references used.
\begin{table}[h]
	\centering
	\begin{tabular}{|c|c|c|c|c|c|c|}
		\hline
		&&&&&&\\[-3mm]
		Material & $m_*$ [a.u.] & Ref. & $\hbar\alpha_\text{R}\,(\text{eV}\text{cm})$ & Ref. & $g_e$ & Ref.\\[1mm]
		\hline
		&&&&&&\\[-3mm]
		InAs & 0.023 & \cite{Ioffe_InAs,Sladek} & $5.71\times10^{-9}$ & \cite{Fan16} & $-15$ & \cite{Hapke02}\\[1mm]
		InSb & 0.014 & \cite{Ioffe_InSb,Sladek} & $3\times10^{-10}$ & \cite{Fan16} & $-51$ & \cite{Litvinenko08}\\[1mm]
		\hline
	\end{tabular}
	\caption{\footnotesize Material parameters used in the simulations: effective electron mass $m_*$, Rashba coefficient $\alpha_\text{R}$, and Land\'{e} g-factor $g_e$, together with their references.}\label{table:material}
\end{table}
\vspace{-8mm}
\paragraph{Energy surfaces in momentum space.}
Motivated by common practice in spintronics, we provide plots of the eigenvalues of the matrix-valued ballistic Hamiltonian~\eqref{eq:ballistic} for each test case in standard momentum representation.
Since $\widehat{H}_{\operatorname{bal}}(p)$ in \eqref{eq:ballistic}-\eqref{eq:nballistic} is a Hermitian $2\times 2$ matrix, its eigenvalues are real and can be shown as two momentum-dependent profiles $\lambda_1(p)$ and $\lambda_2(p)$.
The latter provide useful insight into the dynamics.
In higher-dimensional problems, the eigenvalues would depend parametrically on all momentum components, thereby identifying energy \emph{surfaces} rather than one-dimensional curves.
Therefore, inspired by the analogies between spintronics and chemical physics \cite{BaNi20,StBaDSPiWh16}, we will refer to the energy eigenvalues as \emph{energy surfaces} or \emph{energy surfaces in momentum space}.

\begin{figure}
	\centering
	\includegraphics[width=\textwidth]{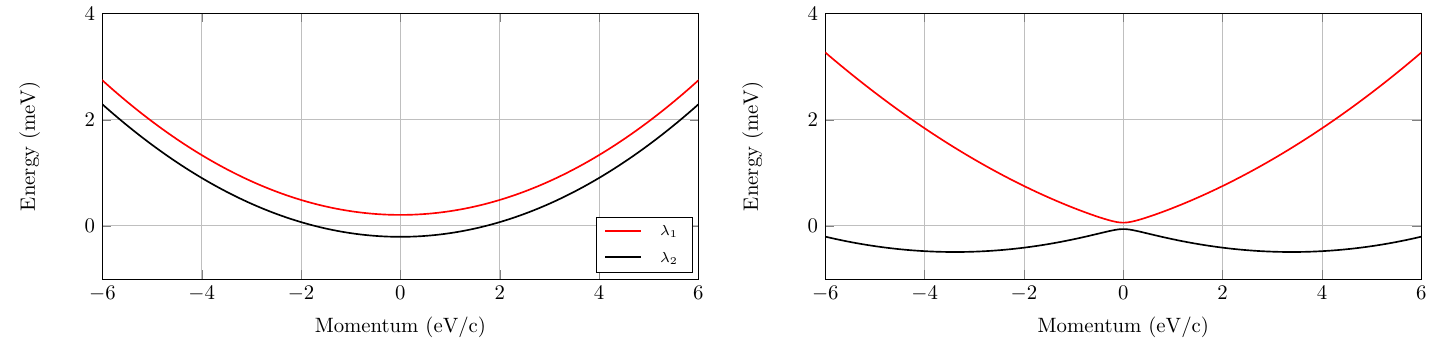}\vspace{-.25cm}
	\caption{\footnotesize
		Energy surfaces in momentum space for the ballistic test cases.
		Left: Zeeman-dominated regime (InSb).
		Right: Rashba-dominated regime (InAs).
		Red: $\lambda_1$; black: $\lambda_2$.
	}\label{1-PES_ballistic}
\end{figure}
\subsection{Ballistic dynamics in the Zeeman-dominated regime}
We begin with a ballistic Hamiltonian and choose material parameters corresponding to InSb.
From Table~\ref{table:material} we obtain the Rashba constant $\hbar\alpha_\text{R}=3\times10^{-10}~\text{eV}\text{cm}$ and the effective electron mass $m=m_*/m_e=0.014$, which give $E_{\text{SO}}=0.83~\mu\text{eV}$.
For the magnetic field we take $B_x=140~\text{mT}$.
With the Land\'{e} g-factor from Table~\ref{table:material}, this yields $E_{\text{Z}}=-206.6~\mu\text{eV}$.
Hence, the coupling parameter in \eqref{eq:defR} is $R=0.008$, so this test case lies well in the Zeeman-dominated regime.
The corresponding energy surfaces are shown in Figure~\ref{1-PES_ballistic}, left panel, and we can see that the two eigenvalue profiles are given by shifted parabolas.

The initial state is the normalized Gaussian wavepacket \eqref{eq:initial_GW} with means $\mu_q=\mu_p=0$ and spatial standard deviation $\sigma_q=\sqrt{\hbar/(2m_*\omega)}\approx 0.46~\mu\text{m}$.
This standard deviation corresponds to the ground state of a reference harmonic oscillator with frequency $f=\omega/(2\pi)\approx 3.1~\text{GHz}$.
This value was chosen to fit within the frequency range of the oscillators studied in the non-ballistic test cases below.
Furthermore, the final simulation time is $T=217.7~\text{ps}$.
This final time is chosen in both ballistic test cases to ensure that the purity of the fully quantum simulation has effectively relaxed to its asymptotic value of $1/2$.

Recall that in all simulations we use $N=500$ koopmons and $\alpha=0.5$.
The MTE simulations use the same number of trajectories in order to compare the two MQC models at the same sampling resolution.
We note that at the same particle number, the koopmon implementation is computationally more expensive than MTE because its backreaction terms require the evaluation of particle interactions.

\begin{figure}[h!]
	\centering
	\includegraphics[width=.85\textwidth]{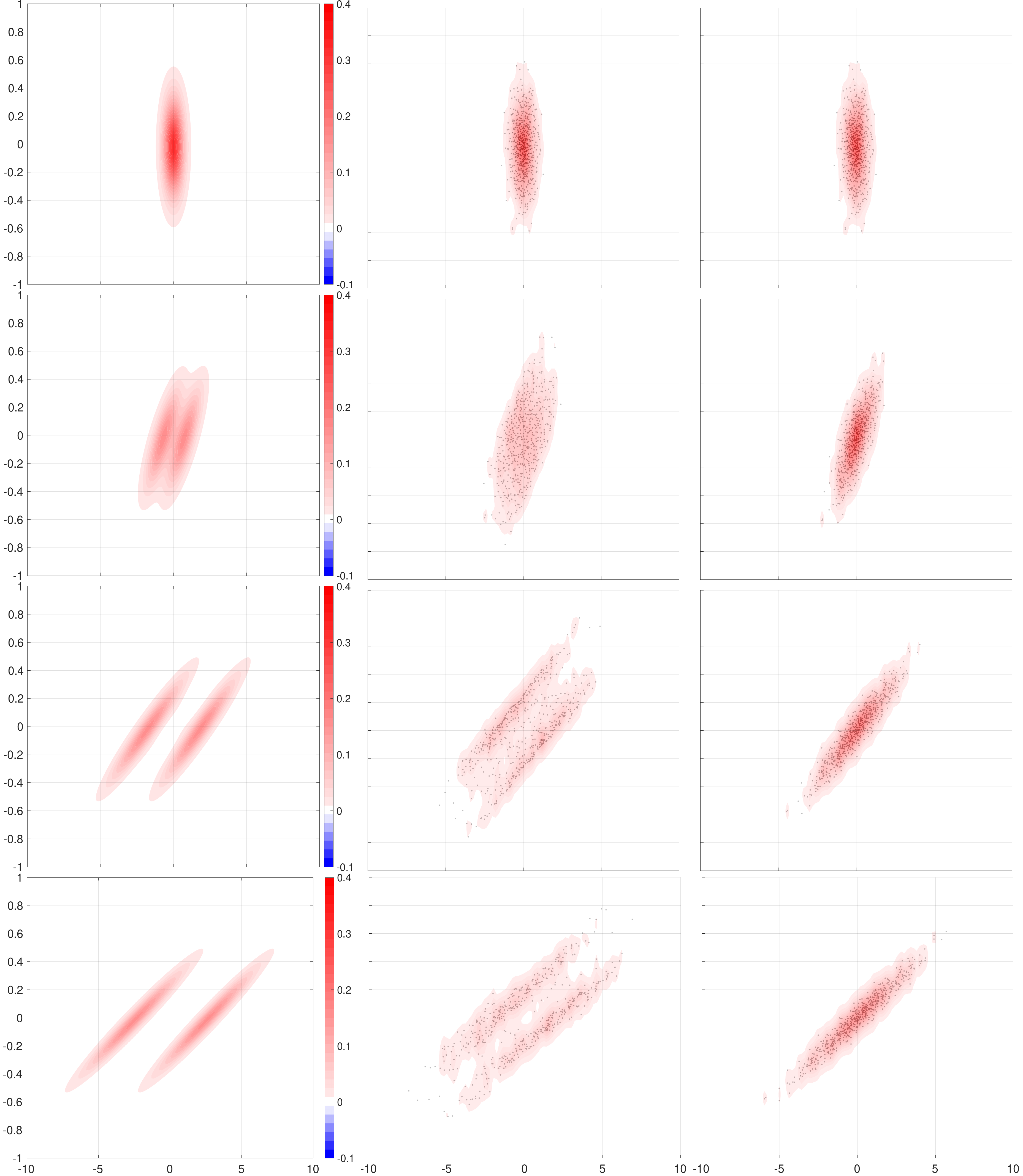}\vspace{-.25cm}
	\caption{\footnotesize
		Time evolution in the classical sector for the ballistic test case in the Zeeman-dominated regime (InSb).
		Columns: Quantum (left), koopmons (middle), Ehrenfest (right).
		Rows: $t=0,62.2,155.5$, and $217.7$~\text{ps}.
		Phase space: position $[q]=\mu\text{m}$ (horizontal) and momentum $[p]=\text{eV}/\text{c}$ (vertical).
		Particle plots with $N=500$ and $\alpha=0.5$ include the smoothed density $D(\bz,t)$.
	}\label{3-bZD-C}
\end{figure}

\paragraph{Classical sector.}
Figure~\ref{3-bZD-C} displays the classical phase-space evolution.
The first column shows the Wigner distribution obtained from the quantum simulation, with the snapshots corresponding to different instants in time.
The second and third columns show the analogous results for the koopmon method and MTE, respectively.

Let us first consider the quantum result.
The top row displays the initial Gaussian distribution.
As time progresses, the packet undergoes three characteristic effects: a clockwise rotation in phase space, a gradual splitting into two branches, and a noticeable squeezing.
At $t=62.2~\text{ps}$, the wavepacket has begun to stretch, but the two emerging branches are still connected.
At later times, the Wigner distribution develops a clear two-peak structure.
These peaks separate further while becoming increasingly narrow.
Integrating the phase-space density over the classical position shows that the momentum marginal density remains essentially Gaussian throughout the evolution.

The koopmons and the MTE method start from the same initial particle cloud, as shown in the first row.
Both methods capture the overall clockwise rotation and the global squeezing of the distribution.
However, the MTE simulation produces a single particle cloud that stays centered at the origin for all times.
In particular, no signature of splitting appears.
In contrast, the koopmon simulation begins to form two distinct regions of higher density.
Although the separation between these two portions is smaller than in the quantum result, the onset of the splitting is clearly visible.
This shows that the koopmons successfully reproduce qualitative features of the orbital quantum dynamics, while MTE struggles even in the Zeeman-dominated regime, where the MQC Rashba coupling is weak.

\begin{figure}
	\centering
	\includegraphics[width=\textwidth]{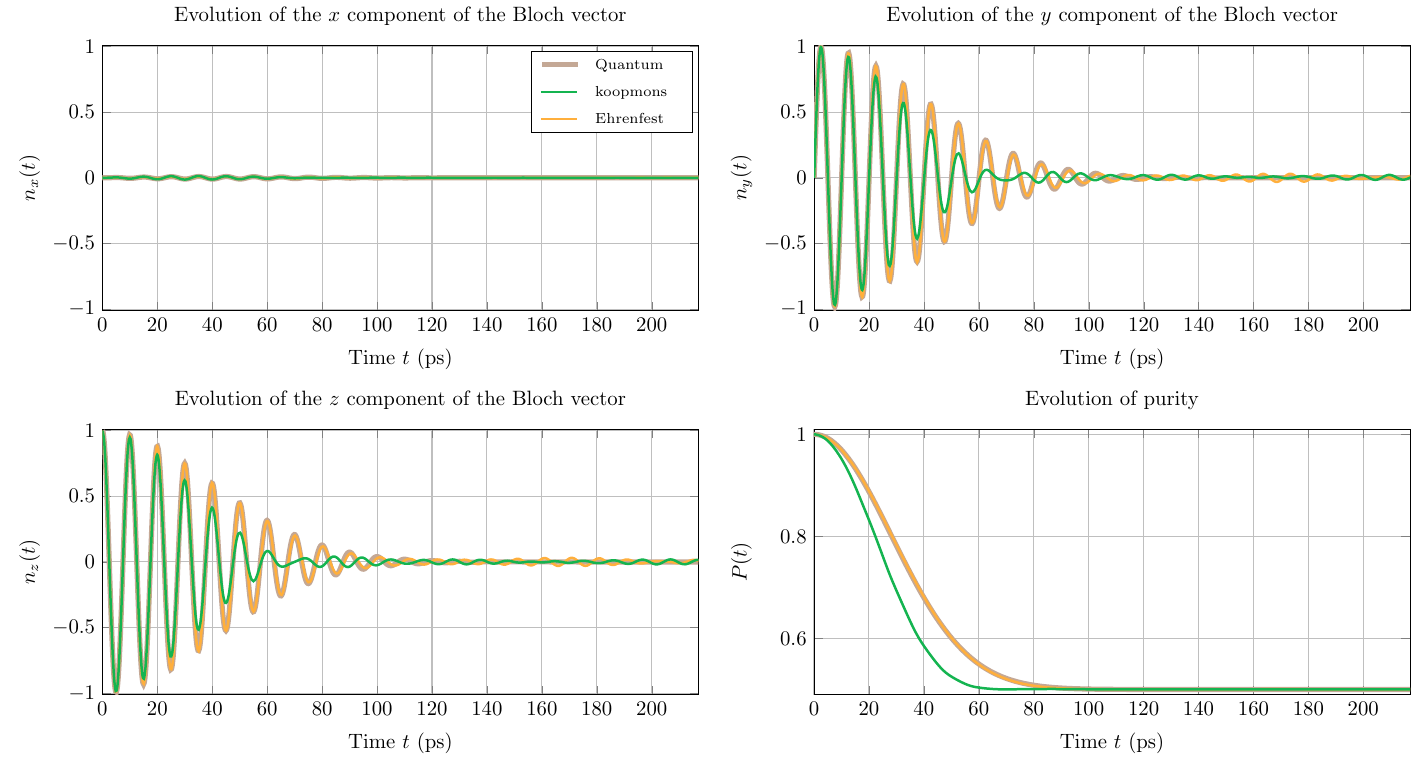}\vspace{-.25cm}
	\caption{\footnotesize
		Time evolution of the Bloch vector components and purity (bottom right) for the ballistic, Zeeman-dominated case (InSb).
	}\label{4-BZD-B}
\end{figure}

\paragraph{Quantum sector and spin-orbit correlations.}
The results for the quantum sector are shown in Figure~\ref{4-BZD-B}.
The two panels in the top row display the time evolution of the $x$- and $y$-components of the three-dimensional Bloch vector $\bn=(n_x,n_y,n_z)^T$.
The panel in the lower left shows the evolution of $n_z$, and the lower-right panel displays the purity.
In all panels, the quantum solution is shown as a brown solid line, the koopmon result as a green solid line, and the MTE result as an orange solid line.

Note that all curves start at $\bn(0)=(0,0,1)$, which reflects our choice of initial state aligned with $\widehat\sigma_z$.
In the $x$-component, the dynamics are essentially trivial: the quantum curve remains almost constant, and both particle methods accurately reproduce this behavior.
The other two components show richer behavior.
Both $n_y(t)$ and $n_z(t)$ exhibit periodic oscillations in the quantum simulation, with a gradual decay of amplitude as time increases.

For short and intermediate times, the MTE method captures both the frequency and the amplitude of the oscillations very well.
Up to roughly $100~\text{ps}$, the MTE curves remain almost indistinguishable from the quantum reference.
The koopmons also match the quantum period, but only during the initial stage of the evolution.
Indeed, up to about $60~\text{ps}$, the oscillations in the koopmon simulation agree well with the quantum and MTE curves, although their amplitude decreases slightly faster.
For times beyond $100~\text{ps}$, both particle methods exhibit small oscillations around zero that are not visible in the quantum result.
The evolution of purity follows a similar pattern with a decay towards the minimum value 0.5.
MTE reproduces the quantum decay of purity remarkably well.
The koopmons show a stronger reduction in purity between $t=0$ and $t\approx 80~\text{ps}$, indicating increased decoherence in this early stage.
After the minimum is reached, all three curves converge and the long-time behavior is identical across all methods.

\begin{figure}
	\centering
	\includegraphics[width=\textwidth]{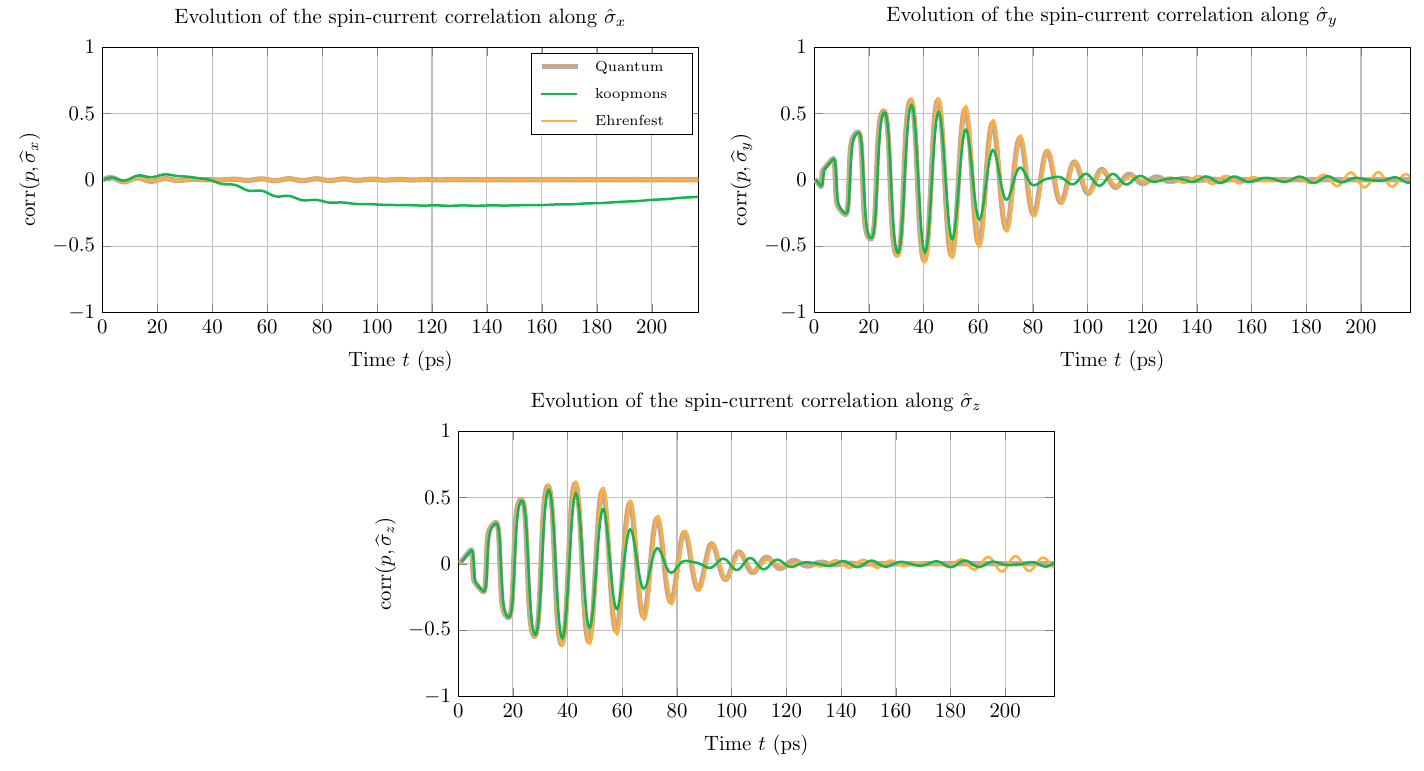}\vspace{-.25cm}
	\caption{\footnotesize
		Time evolution of the spin-momentum correlation components along $\widehat\sigma_x$ (top left), $\widehat\sigma_y$ (top right), and $\widehat\sigma_z$ (bottom) for the ballistic, Zeeman-dominated case (InSb).
	}\label{5-BZD-SC}
\end{figure}

Figure~\ref{5-BZD-SC} shows the evolution of the spin-momentum correlations
\begin{align*}
	\operatorname{corr}(p,\widehat\sigma_k)
	:=\frac{\overline{p\sigma_k}-\bar{p}\,\overline{\sigma}_k}{\sqrt{\overline{p^2}-\overline{p}^2}\sqrt{\overline{\sigma_k^2}-\overline{\sigma}_k^2}},\quad
	k\in\{x,y,z\},
\end{align*}
where we have introduced the bar notation $\bar{A}={\int}\Psi^\dagger\widehat{A}\Psi\,\de q$ for quantum expectation values.
This should not be confused with the regularization notation introduced in Section~\ref{sec:koopmons}.
In the MQC setting, these are related by $\bar{A}={\int}f\langle\widehat{A}\rangle\de q\de p$.

The top-left panel corresponds to $\operatorname{corr}(p,\widehat\sigma_x)$, the top-right panel to $\operatorname{corr}(p,\widehat\sigma_y)$, and the bottom panel to $\operatorname{corr}(p,\widehat\sigma_z)$.
As in the Bloch-vector plots, the $x$-component of the spin-momentum correlations remains zero for all times.
This behavior is matched exactly by the MTE simulation.
The koopmons deviate noticeably and produce small nonzero values.
For the $y$- and $z$-components, the quantum solution again shows periodic oscillations that grow in amplitude until about $40~\text{ps}$, and then decay slowly.
Both MQC particle methods reproduce this qualitative behavior.
All three curves agree well for $t\le 50~\text{ps}$.
After that point, the koopmons exhibit a faster decay of the oscillation amplitude and a shift in the oscillation phase relative to both the quantum and MTE curves.
At late times, both particle methods show once again residual oscillations around zero, which are absent in the fully quantum result.

\subsection{Ballistic dynamics in the Rashba-dominated regime}
In the second test case we again consider a ballistic Hamiltonian, but now using material parameters for InAs.
From Table~\ref{table:material}, the effective electron mass $m=m_*/m_e=0.023$ and Rashba coefficient $\hbar\alpha_\text{R}=5.71\times10^{-9}~\text{eV}\text{cm}$ yield $E_{\text{SO}}=492.1~\mu\text{eV}$.
As in the first test case, we set $B_x=140~\text{mT}$.
With the Land\'{e} g-factor for InAs, this corresponds to $E_{\text{Z}}=-60.8~\mu\text{eV}$.
Hence the coupling ratio in \eqref{eq:defR} is $R=16.2$, which places this system in the Rashba-dominated regime.
The corresponding energy surfaces in momentum space are shown in the right panel of Figure~\ref{1-PES_ballistic}. We observe that the strong Rashba coupling produces a single avoided crossing of the eigenvalue profiles at $p=0$.
As we will see, this strong coupling regime leads to significant quantum interference patterns that tend to take over phase-space dynamics.
In this quantum-dominated dynamics, the use of MQC approaches is clearly a stretch in terms of their regime of validity.
Nevertheless, several qualitative and quantitative features of the fully quantum dynamics are still reproduced with unexpected accuracy.

The initial conditions are identical to those of the first test case, but the final time of the simulation is now $T=46~\text{ps}$.

\begin{figure}[h!]
	\centering
	\includegraphics[width=.85\textwidth]{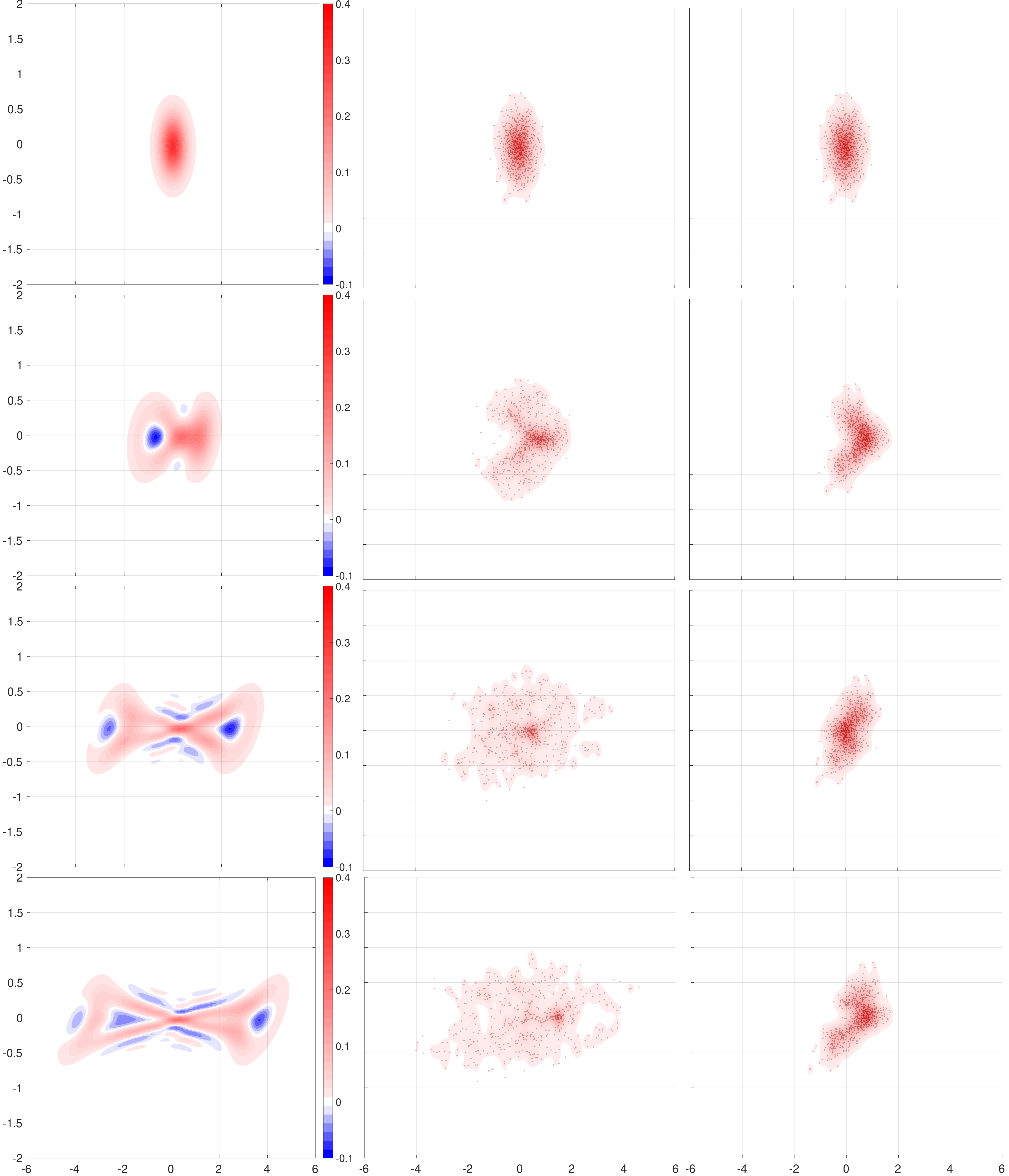}\vspace{-.25cm}
	\caption{\footnotesize
		Time evolution in the classical sector for the ballistic test case in the Rashba-dominated regime (InAs).
		Columns: Quantum (left), koopmons (middle), Ehrenfest (right).
		Rows: $t=0,13.1,32.8$, and $46$~\text{ps}.
		Phase space: position $[q]=\mu\text{m}$ (horizontal) and momentum $[p]=\text{eV}/\text{c}$ (vertical).
		Particle plots with $N=500$ and $\alpha=0.5$ include the smoothed density $D(\bz,t)$.
	}\label{6-BRD-C}
\end{figure}

\paragraph{Classical sector.}
The results for the classical sector are shown in Figure~\ref{6-BRD-C}, following the same illustrative scheme as in the previous case.
The evolution of the Wigner distribution is qualitatively very different from the Zeeman-dominated case shown in Figure~\ref{3-bZD-C}.
The most striking feature is the early and persistent appearance of negative values, which signal strong quantum interference.
At $t=13.1~\text{ps}$, the negativities are mostly concentrated on the left side of the distribution and are surrounded by a ring of positive values, although its maximum appears to shift towards the right.
Then at later times these regions become more delocalized and increasingly fragmented.
The center of the distribution shows alternating strips of positive and negative values -- a clear fingerprint of quantum correlation effects.
Despite these quantum effects, the Wigner distribution gradually splits into two main parts.
One part drifts toward negative $q$, while the other part moves toward positive $q$.
The latter carries most of the positive values, indicated by the deep-red region.

The MTE simulation fails to capture these features.
Although parts of the distribution shift to the right at early times, its overall support remains localized.
For example, in the final snapshot the Wigner distribution covers $p\in[-5,5]$, whereas the MTE particles remain confined to the much smaller momentum interval $[-2,2]$.
The MTE distribution also lacks the broad and symmetric deformation visible in the quantum simulation.

The koopmons behave very differently.
First, the overall extent of the particle distribution matches the quantum result more closely.
Particles spread beyond the domain occupied in the MTE simulation and populate regions consistent with the quantum support.
Second, the deep-red concentration of mass at positive $q$ is reproduced.
Most importantly, the koopmon simulation captures certain structural details that are absent in Ehrenfest dynamics.
At the right boundary in the final snapshot, the Wigner distribution shows a sharp ring structure: a localized negative region surrounded by positive values that is connected to the central mass.
The koopmons cannot reproduce the negative Wigner values themselves, but show a related coarse geometric feature: the particle density is reduced near the center of the ring and enhanced around it.
A similar pattern appears on the left side.
Despite the overall loss in accuracy, this level of qualitative agreement remains noteworthy.
Negative values in the Wigner distribution typically signal a regime in which MQC methods become hardly applicable.
Nevertheless, in this test case the koopmons manage to reproduce several key qualitative structures with surprising accuracy.

\begin{figure}
	\centering
	\includegraphics[width=\textwidth]{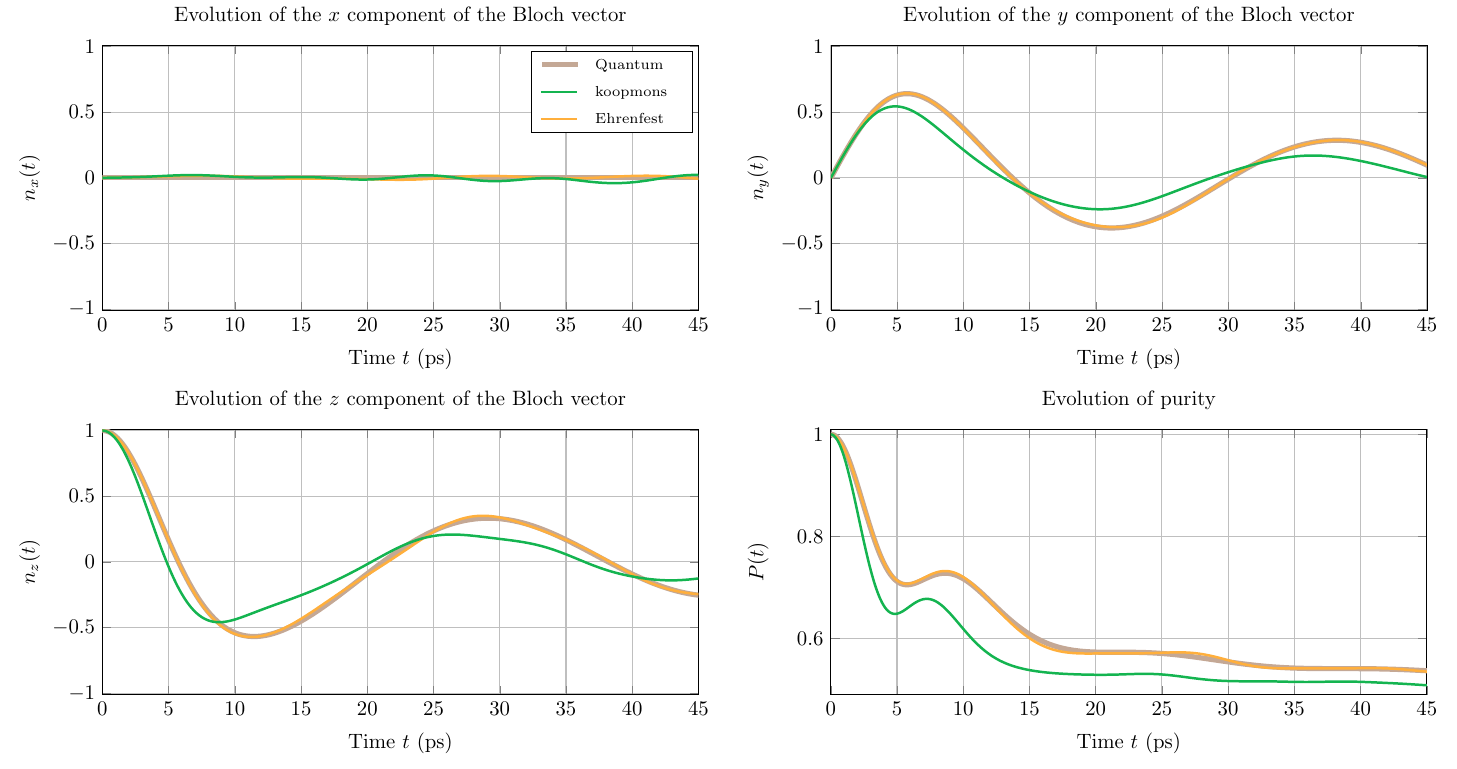}\vspace{-.25cm}
	\caption{\footnotesize
		Time evolution of the Bloch vector components and purity (bottom right) for the ballistic, Rashba-dominated case (InAs).
	}\label{7-BRD-B}
\end{figure}
\paragraph{Quantum sector and spin-orbit correlations.}
The evolution of the Bloch vector and the purity are shown in Figure~\ref{7-BRD-B}.
As in the first test case, only the $y$- and $z$-components of the Bloch vector evolve in time.
The $x$-component remains essentially constant and is reproduced well by both particle methods, aside from small oscillations in the koopmon curve.
The remaining components show oscillatory behavior in the quantum simulation.
The MTE results match these oscillations very closely in both amplitude and phase.
In contrast, the koopmons reproduce the qualitative shape of the oscillations but deviate quantitatively.
The phase shift is visible early on, and the amplitude decays more rapidly.
This discrepancy leads to a more pronounced loss of coherence in the koopmon simulation, as seen in the purity plot.
While the Ehrenfest purity agrees well with the quantum curve, the koopmon purity decreases faster, reflecting the stronger damping observed in the Bloch-vector components.

\begin{figure}
	\centering
	\includegraphics[width=\textwidth]{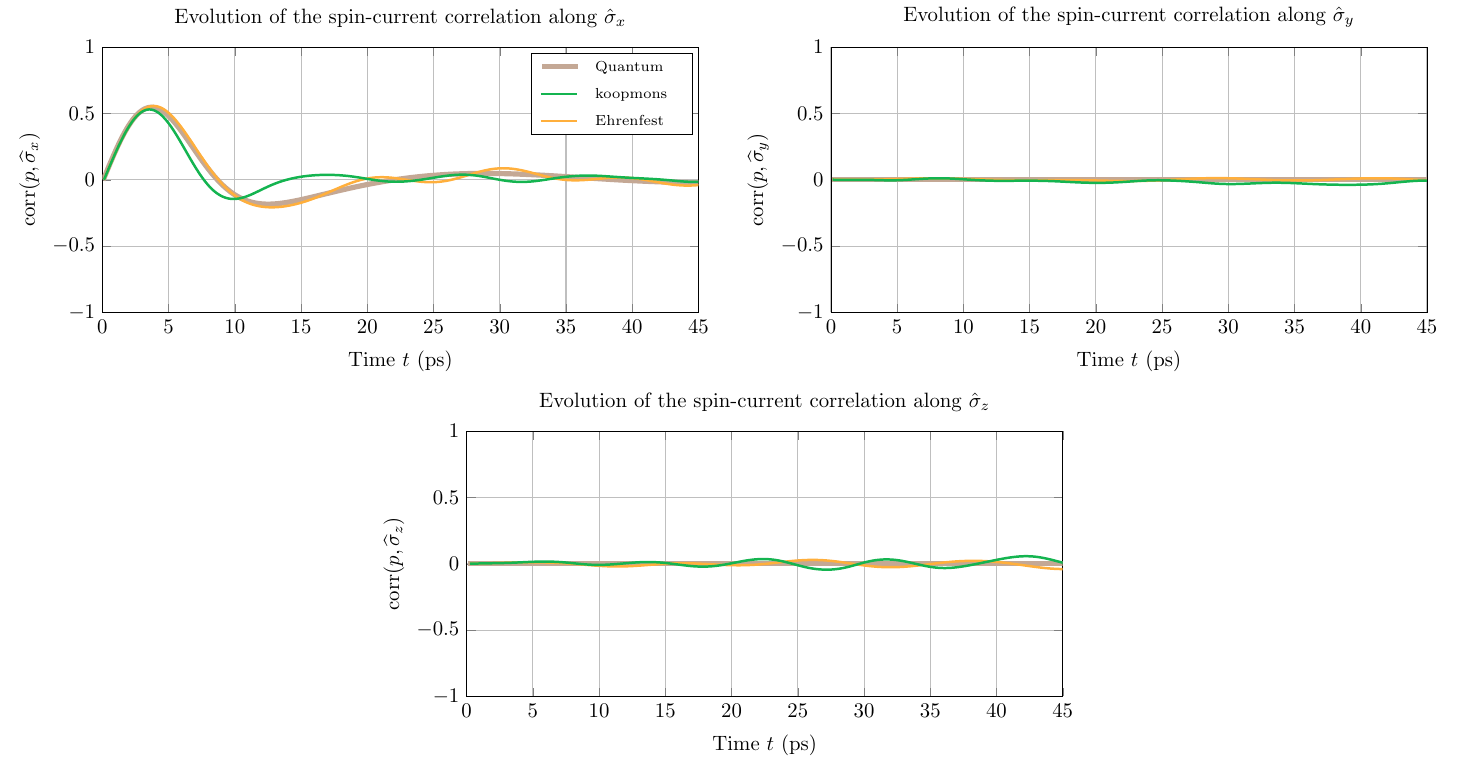}\vspace{-.25cm}
	\caption{\footnotesize
		Time evolution of the spin-momentum correlation components along $\widehat\sigma_x$ (top left), $\widehat\sigma_y$ (top right), and $\widehat\sigma_z$ (bottom) for the ballistic, Rashba-dominated case (InAs).
	}\label{8-BRD-SC}
\end{figure}
The evolution of the spin-momentum correlations is shown in Figure~\ref{8-BRD-SC}.
Here, only the $x$-component is activated.
The MTE method matches the quantum curve again with high accuracy.
The koopmons follow the correct trend but begin to deviate noticeably between $10~\text{ps}$ and $20~\text{ps}$.
In the other two components, we can see small oscillations for the MQC particle methods.
Both methods reproduce the quantum behavior reasonably well, even though the koopmons display slightly larger fluctuations.

\section{Numerical test cases on non-ballistic nanowires}\label{sec:Numerical test cases on non-ballistic nanowires}
In the next three test cases, we study non-ballistic Hamiltonians of the form
\begin{align}\label{eq:nballistic}
	\widehat H(q,p)
	=\widehat{H}_{\operatorname{bal}}(p)+\frac{1}{2}m\omega^2q^2,
\end{align}
which include a HO potential in the classical position $q$, as discussed in Section~\ref{sub:Spin-orbit coupling and Rashba dynamics in nanowire models}.
Such a harmonic potential models a quantum dot, which is frequently encountered in semiconductor nanowires \cite{Fan16,SlSoFl07,SlSoFl06}, and can be used to achieve a spin-orbit qubit in the presence of SOC \cite{Nori}.
More realistic descriptions may rely on a Gaussian potential well instead of the harmonic potential \cite{CaEtAl19}.

As a general feature, non-ballistic dynamics is found to quickly produce significant quantum interference patterns in all coupling regimes, thereby challenging the validity of the MQC approximation.
However, as we will see in this section, several qualitative and quantitative features of the fully quantum dynamics are still reproduced with unexpected accuracy, especially by the koopmon scheme.

\subsection{Non-ballistic dynamics in the Zeeman-dominated regime}\label{sec:nbZd}
For the non-ballistic test cases, one deals with three different energy scales, $E_{\text{SO}}$, $E_{\text{Z}}$, and $E_{\text{HO}}$.
It has proven useful to introduce the dimensionless parameter
\begin{align}\label{def:xi}
	\xi
	:=
	\begin{cases}
		E_{\text{SO}}/E_{\text{HO}}, & \text{if $R>1$ (Rashba-dominated)},\\
		|E_{\text{Z}}|/E_{\text{HO}}, & \text{if $R<1$ (Zeeman-dominated)},
	\end{cases}
\end{align}
where $E_{\text{HO}}=\omega$ denotes the characteristic energy of the HO ($\hbar=1$ in atomic units).
Physically relevant systems often correspond to $\xi\ll 1$ \cite{SlSoFl07,Nori}, meaning that the harmonic potential dominates the dynamics.
However, our numerical experiments indicate that when the HO potential is largely dominant, Rashba effects are less evident and the resulting dynamics are governed primarily by the external confining potential.
Since our main goal is to isolate and study the influence of Rashba coupling, we therefore choose HO frequencies that are lower than those typically encountered in fully realistic setups.
Consequently, all test cases considered here correspond to regimes in which the HO potential remains the leading energy scale, but does not entirely dominate over the Rashba or Zeeman energies.\\

Similar to the first ballistic test case, here we consider InSb parameters.
Recall that for this material we have $\hbar\alpha_\text{R}=3\times10^{-10}~\text{eV}\text{cm}$ and the spin-orbit energy is $E_{\text{SO}}=0.83\,\mu\text{eV}$.
In contrast to the ballistic case, we now use a magnetic field of $B_x=-5.6\,\text{mT}$ which, together with the Land\'{e} g-factor in Table~\ref{table:material}, yields $E_{\text{Z}}=8.3\,\mu\text{eV}$.
Hence, the coupling parameter defined in \eqref{eq:defR} is $R=0.2$, corresponding to the Zeeman-dominated regime.

\begin{figure}
	\centering
	\includegraphics[width=\textwidth]{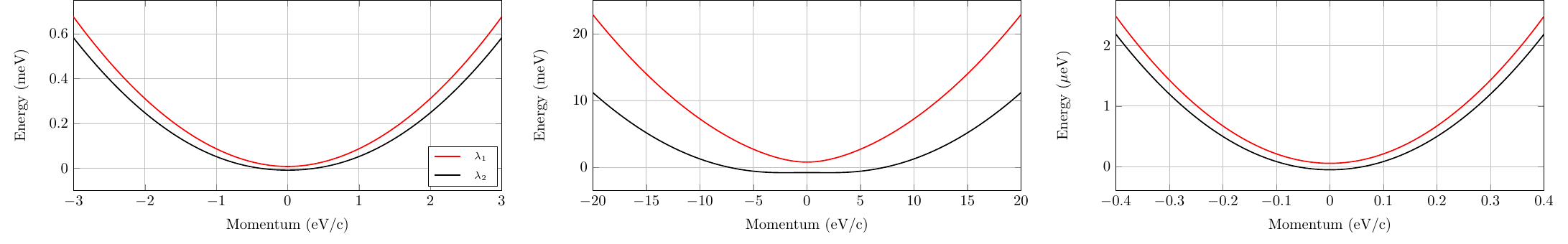}\vspace{-.5cm}
	\caption{\footnotesize
		Energy surfaces in momentum space for the non-ballistic test cases.
		Left: Zeeman-dominated (InSb).
		Middle: Rashba-dominated (InAs).
		Right: Zeeman-dominated (GaAs).
		Red: $\lambda_1$; black: $\lambda_2$.
	}\label{2-PES_nballistic}
\end{figure}

For the characteristic energy of the HO we choose $\hbar\omega=16.6\,\mu\text{eV}$ ($\approx 4\,\text{GHz}$), which, according to \eqref{def:xi}, yields $\xi=0.5$.
As anticipated, this value indicates that the HO potential remains the dominant energy scale relative to the Zeeman energy.
The corresponding momentum energy surface for the associated ballistic Hamiltonian $\widehat{H}_{\operatorname{bal}}$ in \eqref{eq:nballistic} is shown in the left panel of Figure~\ref{2-PES_nballistic}.
As can be seen, the two eigenvalue profiles appear as shifted parabolas; however, they are not parallel and instead tend to approach each other near the origin.

The initial conditions are given by $\mu_q=0$, an initial momentum kick of $\mu_p=4.5\,\text{eV}/\text{c}$, and a variance $\sigma_q^2$ corresponding to the ground state of the HO associated with the chosen frequency $\omega$.
The final time of the simulation was set to $T=822.4\,\text{ps}$.
We note that, in all three non-ballistic test cases, the final simulation times were chosen \emph{a posteriori} based on the onset of pronounced negativities in the Wigner distribution of the fully quantum solution.
As discussed earlier, these negativities are a characteristic signature of genuine quantum correlation effects.
Once they become dominant, they typically lead to an increased discrepancy between the purity obtained from the MQC particle methods and that of the fully quantum simulation, thereby marking a natural endpoint for the MQC simulations.

\begin{figure}[h!]
	\centering
	\includegraphics[width=.85\textwidth]{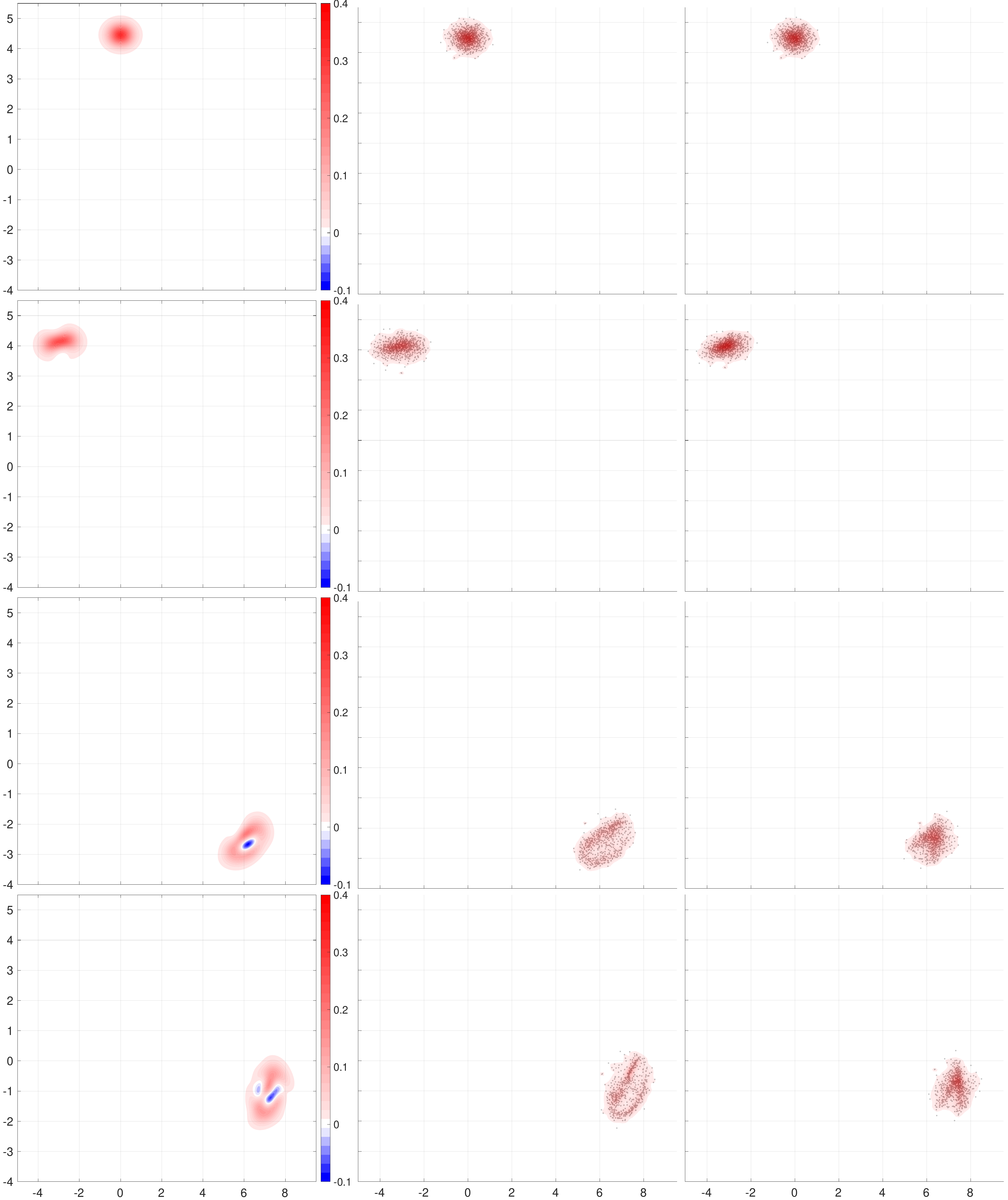}\vspace{-.25cm}
	\caption{\footnotesize
		Time evolution in the classical sector for the non-ballistic test case in the Zeeman-dominated regime (InSb).
		Columns: Quantum (left), koopmons (middle), Ehrenfest (right).
		Rows: $t=0,235, 587.4$, and $822.4$~\text{ps}.
		Phase space: position $[q]=\mu\text{m}$ (horizontal) and momentum $[p]=\text{eV}/\text{c}$ (vertical).
		Particle plots with $N=500$ and $\alpha=0.5$ include the smoothed density $D(\bz,t)$.
	}\label{12-NBZD-C}
\end{figure}
\paragraph{Classical sector.}
The results for the classical sector are shown in Figure~\ref{12-NBZD-C}.
The Wigner distribution (first column) exhibits a clockwise motion in phase space, accompanied by an intricate deformation of the wavepacket.
The last two snapshots show the formation of one local minimum at $t=587.4\,\text{ps}$, followed by an additional one at $t=822.4\,\text{ps}$, both visible as blue regions.
These minima are surrounded by regions of positive distribution.

Other than the global clockwise motion, the koopmon results qualitatively reproduce finer structural features that are absent in the Ehrenfest dynamics.
The contrast is already visible at $t=587.4\,\text{ps}$: the koopmons place more particles near the boundary of the distribution and fewer in the interior, reflecting the central minimum in the Wigner distribution.
The Ehrenfest solution (third column), in contrast, remains more concentrated in the middle and does neither exhibit the same squeezing nor the thinning of particles at the center.
This discrepancy becomes even clearer at the final time.
While both particle methods fail to reproduce the fine details, the overall shape of the distribution -- its extent, its deformation, and its reduced density in the interior -- is captured significantly better by the koopmons.
To verify that the visible differences in the MTE result are not primarily caused by insufficient particle sampling, we repeated this MTE test case with $N=5000$ trajectories.
Compared with the $N=500$ MTE simulation, the maximum absolute differences over the full time interval were $3.97\times10^{-3}$ for the purity, $6.74\times10^{-3}$ for $\langle q\rangle$, $1.20\times10^{-2}$ for $\langle p\rangle$, and $2.25\times10^{-2}$ for $\operatorname{Var}(q)$.
The corresponding relative differences for $\langle q\rangle$, $\langle p\rangle$, and $\operatorname{Var}(q)$ were $5.17\times10^{-4}$, $9.14\times10^{-4}$, and $1.74\times10^{-2}$, respectively.
Thus, increasing the number of MTE trajectories by one order of magnitude changes these observables only mildly and does not remove the discrepancy with the fully quantum dynamics.

\begin{figure}
	\centering
	\includegraphics[width=\textwidth]{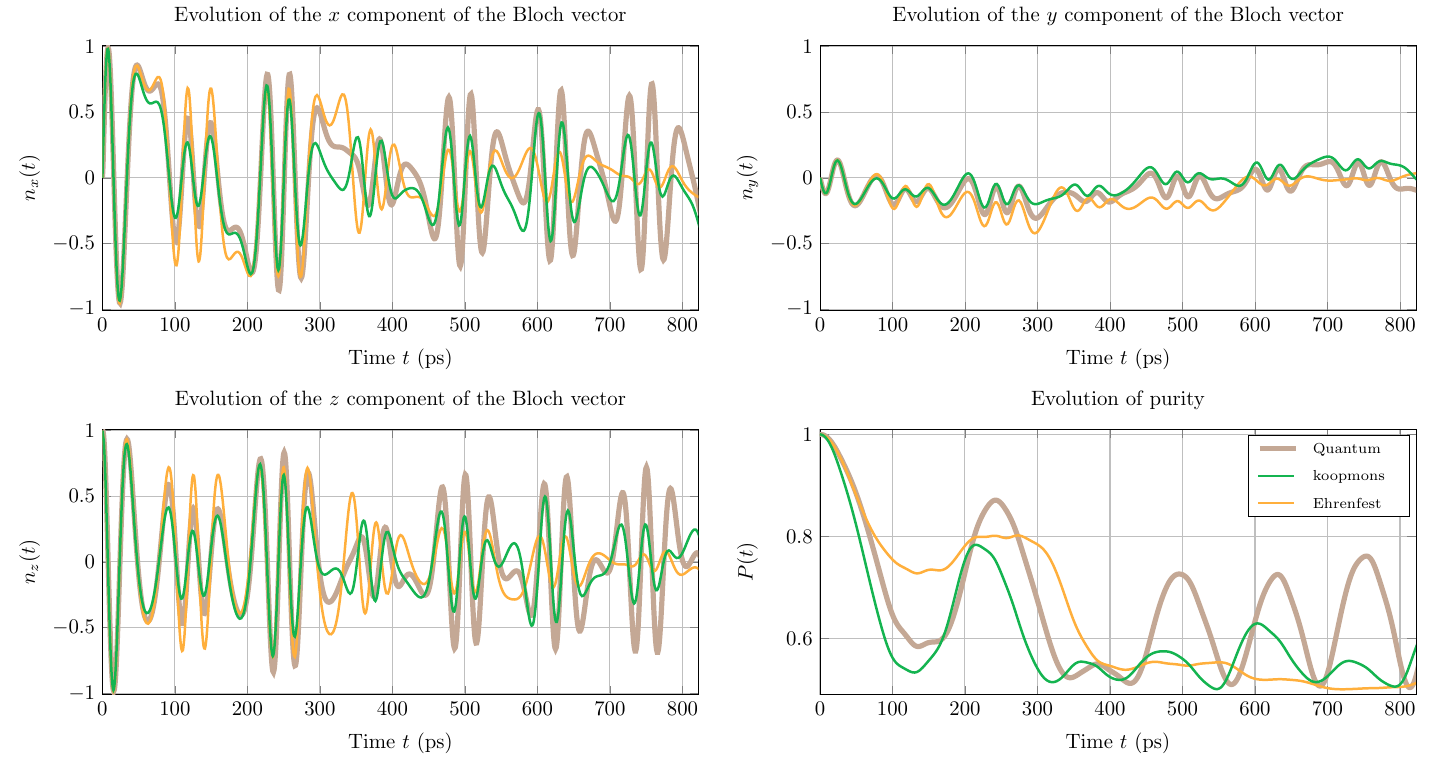}\vspace{-.25cm}
	\caption{\footnotesize
		Time evolution of the Bloch vector components and purity (bottom right) for the non-ballistic, Zeeman-dominated case (InSb).
	}\label{13-NBZD-B}
\end{figure}
\paragraph{Quantum sector and spin-orbit correlations.}
The evolution of the Bloch vector and the purity are shown in Figure~\ref{13-NBZD-B}.
The quantum result shows pronounced oscillations in all three Bloch components.
The amplitudes are large in the first (top left) and third (bottom left) components, and significantly smaller in the second component (top right).
During the first $50\,\text{ps}$, both the koopmons and the MTE dynamics agree very well with the quantum solution across all components.
For later times up to approximately $300\,\text{ps}$, both particle methods begin to deviate in amplitude, although the oscillation periods are still captured reasonably well.
These deviations are most visible in the first and third components: the koopmons slightly underestimate the amplitudes, whereas the MTE method slightly overestimates them.
A different behavior appears for $t\in[300,450]\,\text{ps}$.
In this transient regime, neither particle method matches the quantum solution satisfactorily in any component.
Afterwards, however, the koopmons regain accuracy.
For later times they again track the quantum behavior much more closely, reproducing both the period and the amplitude of the oscillations.
The purity plot reflects these observations directly.
The koopmons show the correct number of minima and maxima -- six minima and five maxima -- although with a slight shift in time.
In contrast, the Ehrenfest method deviates much earlier and begins to drift toward the final value close to $1/2$ after roughly $400\,\text{ps}$.

\begin{figure}
	\centering
	\includegraphics[width=\textwidth]{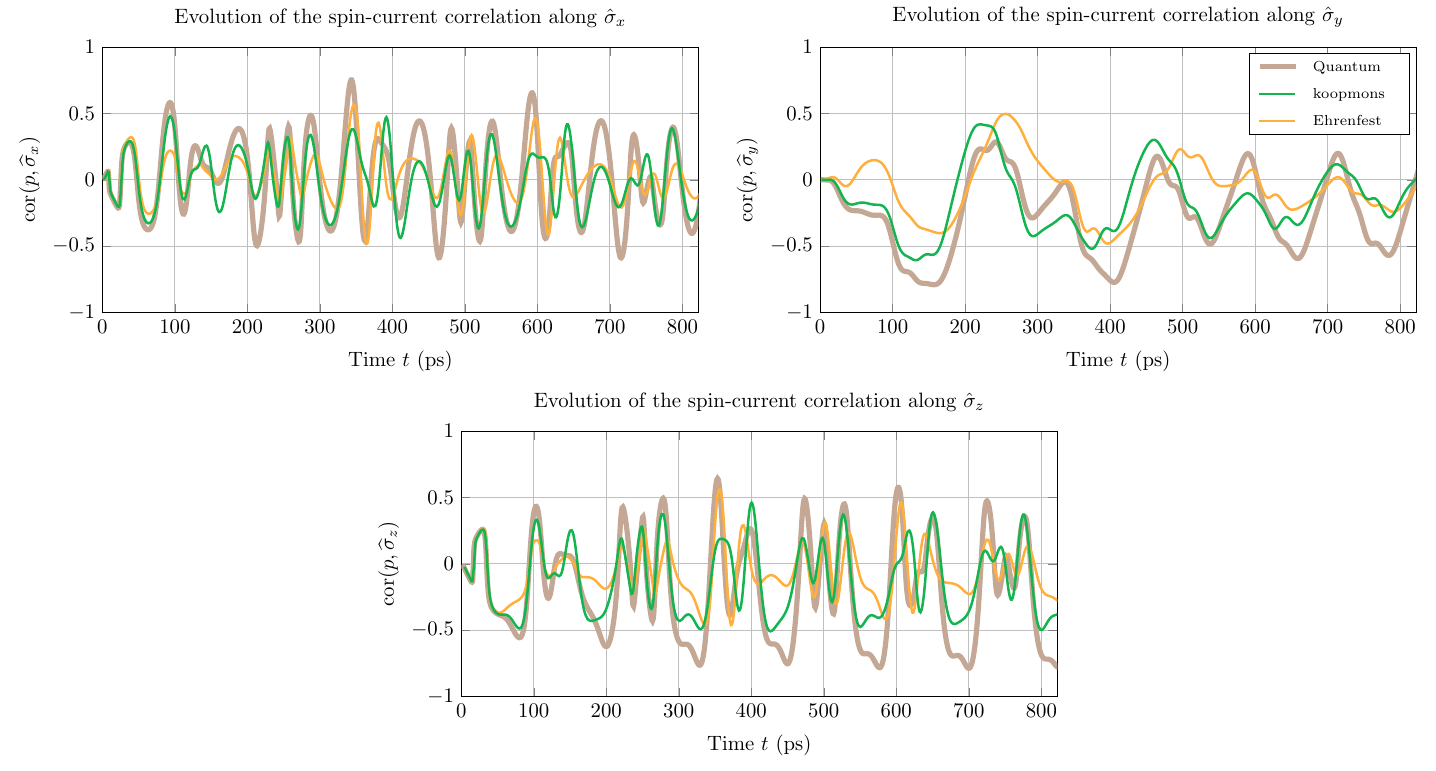}\vspace{-.25cm}
	\caption{\footnotesize
		Time evolution of the spin-momentum correlation components along $\widehat\sigma_x$ (top left), $\widehat\sigma_y$ (top right), and $\widehat\sigma_z$ (bottom) for the non-ballistic, Zeeman-dominated case (InSb).
	}\label{14-NBZD-SC}
\end{figure}
The evolution of the spin-momentum correlations is shown in Figure~\ref{14-NBZD-SC}.
During the first $50\,\text{ps}$, both particle methods agree well with the quantum solution in the first (top left) and third (bottom) components.
In the second component (top right), however, only the koopmons reproduce the qualitative behavior of the quantum result.
Although neither particle method is fully satisfactory in the second component, the koopmons show consistently better agreement in both amplitude and period.
This difference becomes especially pronounced for $t\ge 500\,\text{ps}$.
Here the MTE method fails to reproduce any of the peaks or oscillation periods in the second component, whereas the koopmons continue to reflect the main qualitative features.
In the first and third components, the koopmons capture both amplitudes and periods noticeably better throughout the entire simulation.

\subsection{Non-ballistic dynamics in the Rashba-dominated regime}
Similar to the second ballistic test case, here we consider InAs parameters with Rashba coefficient $\hbar\alpha_\text{R}=5.71\times10^{-9}~\text{eV}\text{cm}$ and spin-orbit energy $E_{\text{SO}}=492.1\,\mu\text{eV}$.
In contrast to the ballistic case, we now apply a much stronger magnetic field, $B_x=-1.8\,\text{T}$, which yields $E_{\text{Z}}=781.4\,\mu\text{eV}$.
Hence, the coupling parameter defined in \eqref{eq:defR} becomes $R=1.26$, placing this test case in the Rashba-dominated regime.
In this case, quantum interference largely dominates the fully quantum evolution, taking the validity of the MQC approximation to the very boundary of its applicability.
This reflects especially in the phase-space dynamics, where the discrepancies are manifest in both the Koopman and Ehrenfest models.
Surprisingly enough, however, MQC approaches are shown to capture essential qualitative features of the spin evolution, with the koopmons possessing better accuracy.

The characteristic energy of the HO, $\hbar\omega=737.1\,\mu\text{eV}$ ($\approx 178\,\text{GHz}$), is also chosen significantly larger than in the previous test case, and the value in \eqref{def:xi} is $\xi=0.67$.
As anticipated, this value indicates that the HO energy dominates over the Rashba energy.
The corresponding momentum energy surfaces are shown in the middle of Figure~\ref{2-PES_nballistic}.
As visible from the plot, the two eigenvalue profiles approach each other near the origin and separate for large values of $|p|$, resembling the structure of a single avoided crossing of the eigenvalues.

The initial conditions are $\mu_q=0$, an initial kick of $\mu_p=37.3\,\text{eV}/\text{c}$, and a variance $\sigma_q^2$ corresponding to the ground state of the HO associated with the chosen frequency $\omega$.
The final simulation time is $T=5.08\,\text{ps}$.
After this time, the purely quantum effects become dominant and the MQC approach becomes inapplicable.

\begin{figure}[h!]
	\centering
	\includegraphics[width=.85\textwidth]{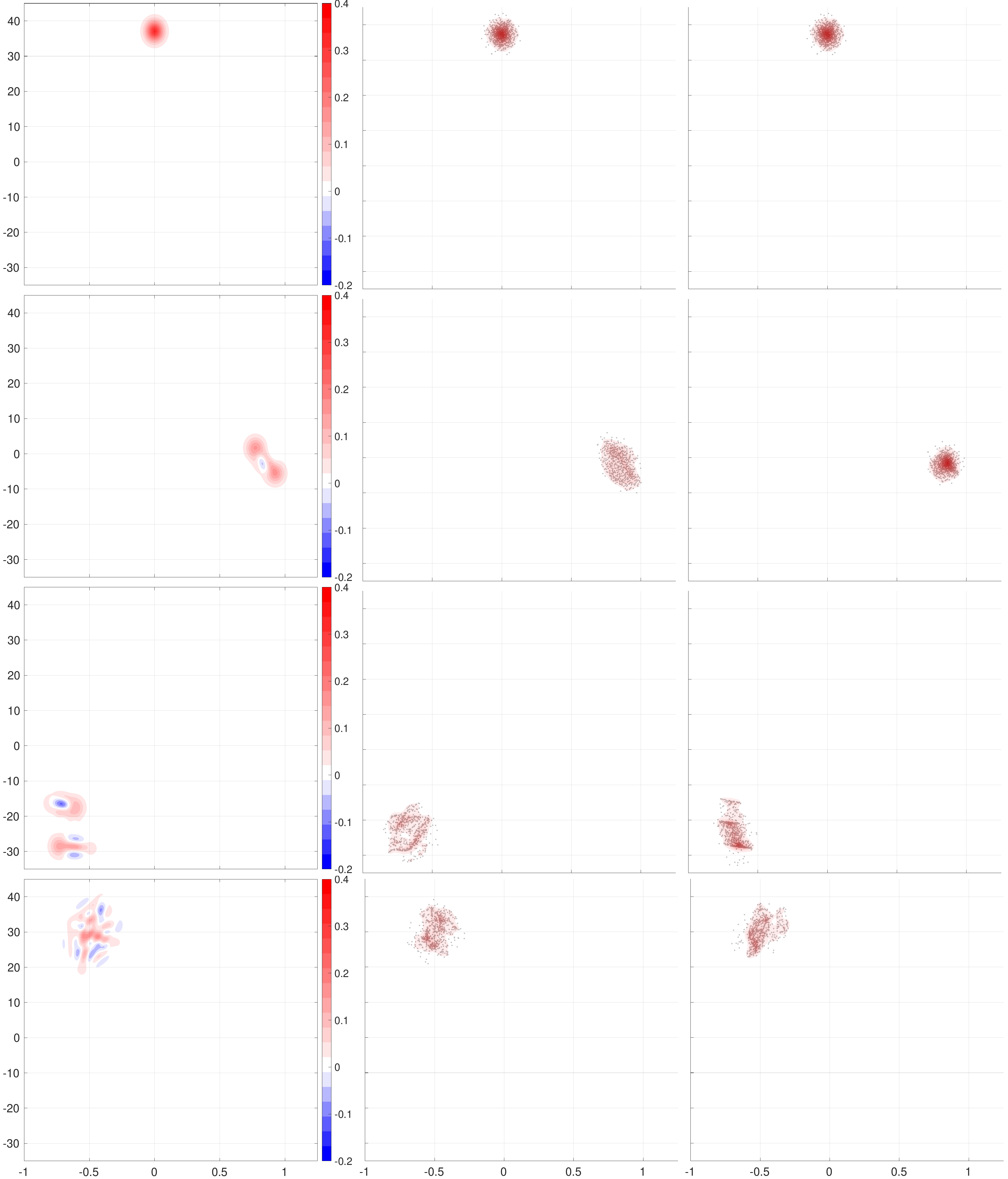}\vspace{-.25cm}
	\caption{\footnotesize
		Time evolution in the classical sector for the non-ballistic test case in the Rashba-dominated regime (InAs).
		Columns: Quantum (left), koopmons (middle), Ehrenfest (right).
		Rows: $t=0,1.45,3.63$, and $5.08$~\text{ps}.
		Phase space: position $[q]=\mu\text{m}$ (horizontal) and momentum $[p]=\text{eV}/\text{c}$ (vertical).
		Particle plots with $N=500$ and $\alpha=0.5$ include the smoothed density $D(\bz,t)$.
	}\label{15-NBRD-C}
\end{figure}
\paragraph{Classical sector.}
The classical-sector results are presented in Figure~\ref{15-NBRD-C}.
We begin with the Wigner distribution (first column).
As in the previous non-ballistic test case, the distribution performs a clockwise motion in phase space.
However, its deformation follows a quite different pattern.
In the second snapshot, the distribution develops two pronounced peaks of positive mass (dark red), together with a negative-valued minimum right in between (light blue).
By $t=3.6\,\text{ps}$, the wavepacket shows a separation into two distinct parts, each showing localized negative regions whose magnitude and area have grown compared to the second snapshot.
The final snapshot shows that these two pieces are recombining, producing a highly fragmented structure dominated by interference effects throughout the support.
Given this complexity, it is not surprising that neither particle-based method is able to reproduce the fine structures of the quantum solution at these times.
Both the koopmons and the Ehrenfest dynamics, however, do reproduce the correct mean values of position and momentum.
Furthermore, in the second snapshot the koopmons capture the elongated shape of the Wigner distribution more accurately.
As we will see shortly, this increased accuracy for small times $t\le 2.5\,\text{ps}$ will also appear for the koopmons in the quantum sector.

\begin{figure}
	\centering
	\includegraphics[width=\textwidth]{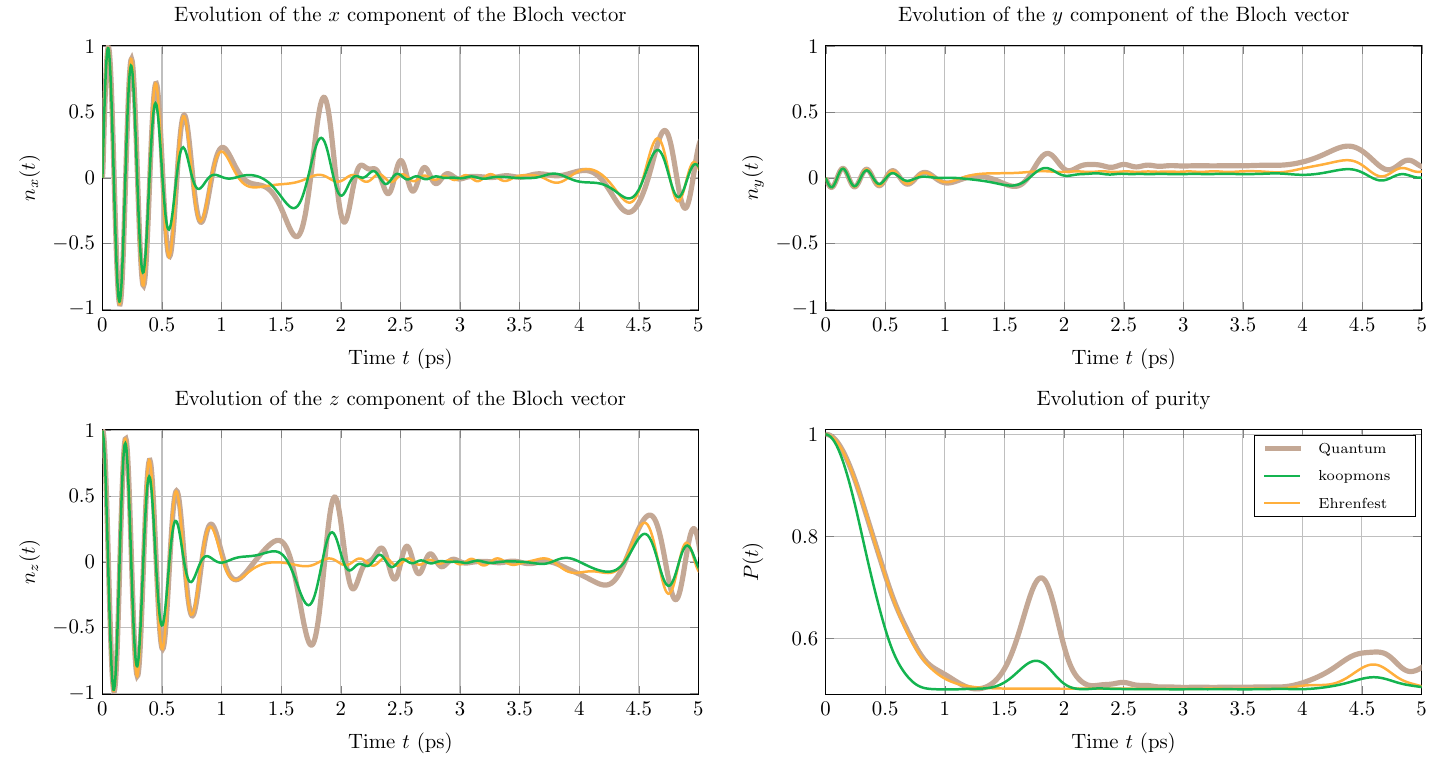}\vspace{-.25cm}
	\caption{\footnotesize
		Time evolution of the Bloch vector components and purity (bottom right) for the non-ballistic, Rashba-dominated case (InAs).
	}\label{16-NBRD-B}
\end{figure}
\paragraph{Quantum sector and spin-orbit correlations.}
Figure~\ref{16-NBRD-B} shows the evolution of the Bloch vector and the purity.
The Ehrenfest dynamics reproduce the quantum result with very high accuracy in all three components of the Bloch vector up to $t\le 1\,\text{ps}$, and this is reflected in the purity plot.
The koopmon simulation, however, produces slightly smaller amplitudes in the first and third components over this early interval, which leads to lower purity levels.
We also observe two prominent amplitude peaks in the first and third components at approximately $t\approx 1.75\,\text{ps}$ and $t\approx 4.75\,\text{ps}$.
A direct comparison with Figure~\ref{15-NBRD-C} indicates that these times coincide with the moments when the two separated parts of the Wigner distribution recombine.
As we can see, only the koopmons capture these peaks.
In particular, both the oscillation periods and the amplitudes are reproduced quite well on the interval $t\in[1.5,2]\,\text{ps}$.
As a consequence, only the koopmons show a corresponding peak in the purity curve, whereas the Ehrenfest simulation remains flat during this time.
For the second peak around $t\approx 4.75\,\text{ps}$, both particle methods reproduce the oscillations, even though small time shifts are visible for both.

\begin{figure}
	\centering
	\includegraphics[width=\textwidth]{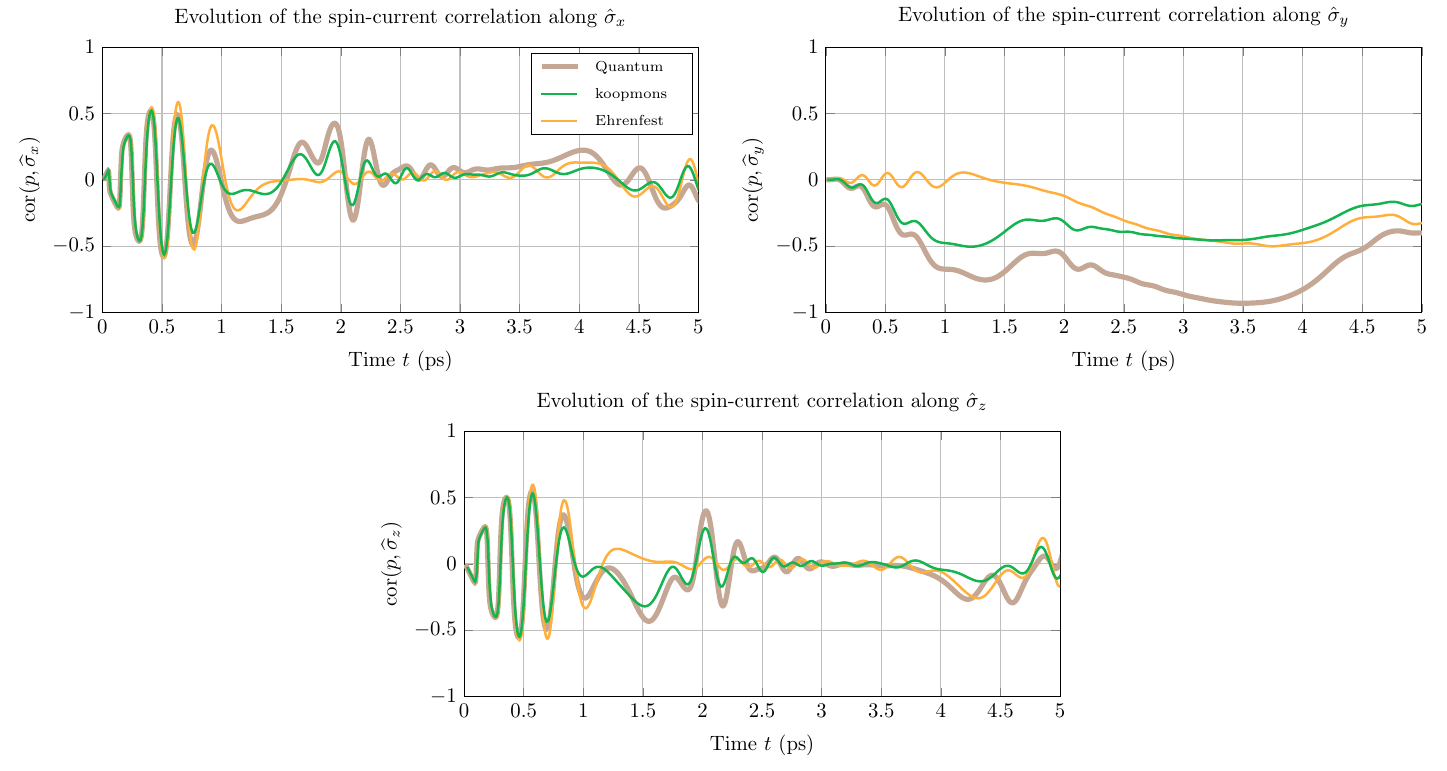}\vspace{-.25cm}
	\caption{\footnotesize
		Time evolution of the spin-momentum correlation components along $\widehat\sigma_x$ (top left), $\widehat\sigma_y$ (top right), and $\widehat\sigma_z$ (bottom) for the non-ballistic, Rashba-dominated case (InAs).
	}\label{17-NBRD-SC}
\end{figure}

The evolution of the spin-momentum correlations, shown in Figure~\ref{17-NBRD-SC}, leads to a very similar set of observations.
For $t\le 1\,\text{ps}$, the koopmon amplitudes match the quantum results even better than in the corresponding Bloch-vector components.
The most significant discrepancy appears in the second component (top right), where the koopmons continue to follow the quantum oscillations over a substantially longer time, while the Ehrenfest dynamics begin to deviate almost immediately.
In the interval $t\in[1.5,2]\,\text{ps}$, the koopmons again reproduce both the amplitudes and periods of the oscillations much better than MTE, whose values remain close to zero.
After a time window in which both particle methods show difficulties tracking the quantum solution, they regain accuracy for $t\ge 4.5\,\text{ps}$, producing nearly identical results thereafter.

\subsection{Appearance of cat-like states in gallium arsenide wires}
In the previous test cases, we considered the materials InSb and InAs.
Owing to its relatively small Rashba coupling parameter, InSb was used to investigate Zeeman-dominated regimes, whereas the larger Rashba coupling in InAs enabled the study of Rashba-dominated dynamics.
Here, we present an additional test case in which the non-ballistic dynamics of a \emph{gallium arsenide} (GaAs) wire exhibits the formation of cat-like structures in phase-space.
Characterized by a particularly low Rashba coefficient, GaAs is an ideal material for studying the Zeeman-dominated dynamics considered in this section.
Furthermore, the appearance of Schr\"odinger cat states makes the present test case particularly challenging for MQC models, which can hardly capture the strong quantum correlations giving rise to the typical interference patterns of cat-state dynamics.

In the following, we consider GaAs with material parameters taken from \cite{Fan16}: Effective electron mass $m=m_*/m_e=0.067$, $\hbar\alpha_\text{R}=0.68\times10^{-11}~\text{eV}\text{cm}$, and $g_e=-0.44$.
The first two parameters yield a spin-orbit energy of $E_{\text{SO}}=2.0~\text{neV}$, while the latter, together with the applied magnetic field $B_x=-4.2~\text{mT}$, results in a Zeeman energy of $E_{\text{Z}}=53.5~\text{neV}$.
Consequently, the coupling parameter in \eqref{eq:defR} is $R=0.076$, which places this test case in the Zeeman-dominated regime.
\begin{figure}[h!]
	\centering
	\includegraphics[width=.85\textwidth]{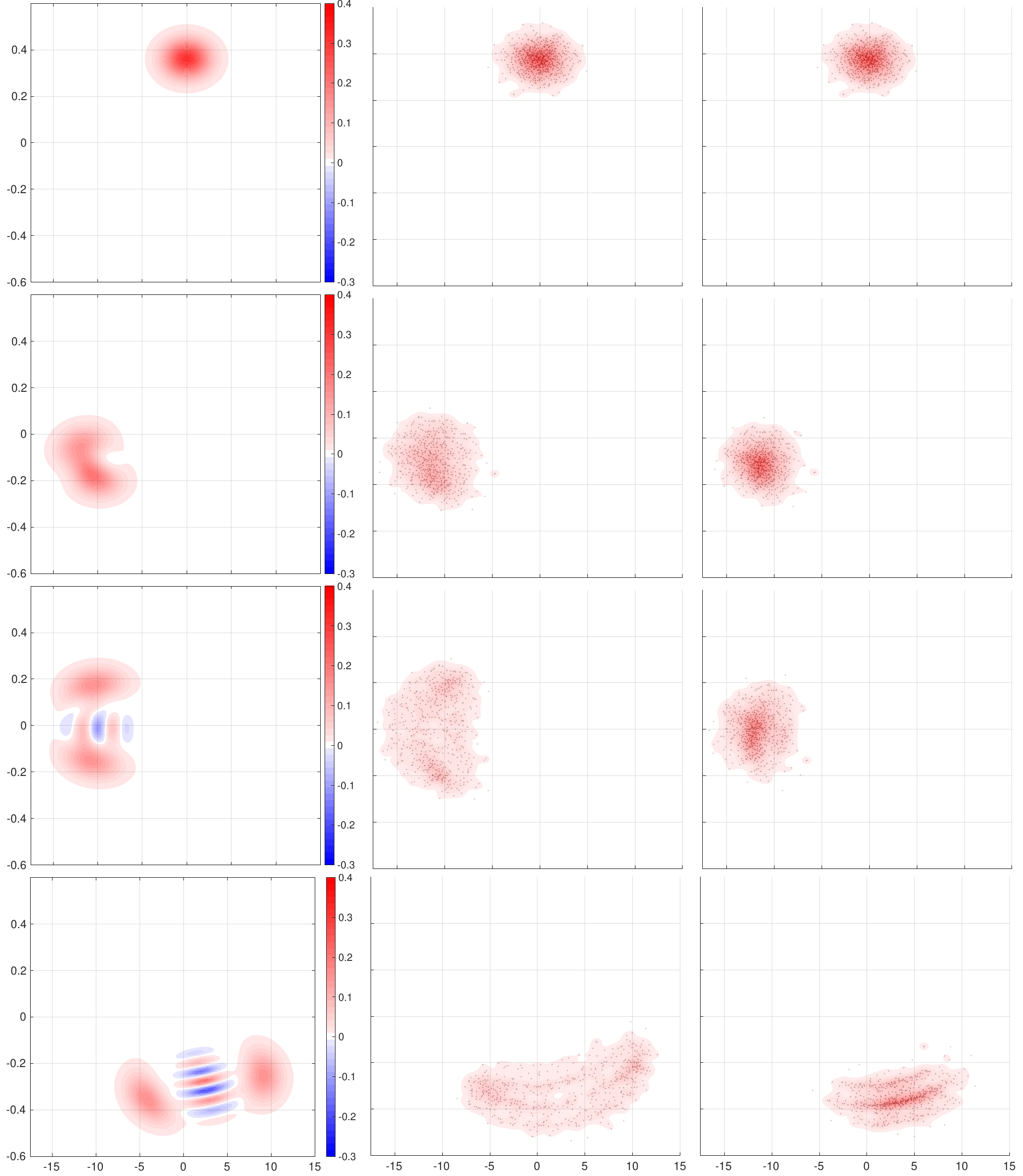}\vspace{-.25cm}
	\caption{\footnotesize
		Time evolution in the classical sector for the non-ballistic test case in the Zeeman-dominated regime (GaAs).
		Columns: Quantum (left), koopmons (middle), Ehrenfest (right).
		Rows: $t=0,61.5,153.8$, and $215.3$~\text{ns}.
		Phase space: position $[q]=\mu\text{m}$ (horizontal) and momentum $[p]=\text{eV}/\text{c}$ (vertical).
		Particle plots with $N=500$ and $\alpha=0.5$ include the smoothed density $D(\bz,t)$.
	}\label{9-NBZD-C}
\end{figure}

The characteristic energy of the HO is chosen as $\hbar\omega=182~\text{neV}$ ($\approx 44\,\text{MHz}$), which yields $\xi=0.29$ according to \eqref{def:xi}.
This frequency is about ninety times lower than the value considered in Section~\ref{sec:nbZd} and much lower than the realistic values modeling semiconductor quantum dots.
Nevertheless, this particular configuration is made interesting by the appearance of cat states that challenge the applicability of the MQC models.
For this test case, the energy surfaces in momentum space are shown in the right panel of Figure~\ref{2-PES_nballistic}.
As in the ballistic Zeeman-dominated case, the two surfaces form vertically shifted parabolas.
The initial conditions are given by $\mu_q=0$, a small initial kick of $\mu_p=0.37~\text{eV}/\text{c}$, and a variance $\sigma_q^2$ corresponding to the ground state of an oscillator with frequency $\omega$.
The final simulation time is $T=215.3~\text{ns}$.

\paragraph{Classical sector.}
Figure~\ref{9-NBZD-C} shows the evolution of the phase-space distributions for the orbital dynamics.
The Wigner distribution evolves along a circular path in phase space -- a direct consequence of the harmonic confinement.
From the initial Gaussian profile, the wavepacket begins to stretch tangentially as it moves along this orbit.
As time increases, a clear bimodal structure emerges.
At $t=153.8~\text{ns}$, the distribution has already developed two pronounced lobes which continue to separate, and by $t=215.3~\text{ns}$ they form two well-defined density peaks.
The region between them no longer exhibits a simple deformation of a Gaussian: instead, it contains a highly structured interference pattern, with oscillatory fringes of alternating sign.
In particular, the negative regions form thin arcs aligned with the circular geometry of the motion, emphasizing the genuinely quantum character of the state.
Overall, the phase-space picture at the final time is strongly reminiscent of a cat state, i.e., a linear superposition of two coherent-state wavepackets linked by a central interference region.

The Ehrenfest simulation captures the overall rotation but fails to reproduce several essential features.
Most of the mass remains confined near the center, even at late times.
Although the distribution begins to stretch in the last snapshot, it does not develop two distinct peaks.
Moreover, the spatial support of the Ehrenfest density is noticeably smaller than that of the Wigner distribution.
For example, at $t=153.8~\text{ns}$, the quantum support lies roughly inside the phase-space box $(q,p)\in[-15,-5]\times [-0.15,0.15]$, while the Ehrenfest distribution occupies a much narrower region.

The koopmon simulation better reflects the quantum result.
First, the spatial extent of the distribution is in much closer agreement with the Wigner distribution.
The koopmons populate the same broad region in phase space and do not remain confined as in the Ehrenfest case.
Second, the koopmons also produce two clear density peaks, with a noticeably reduced number of particles in the central interference region.
This contrasts sharply with the MTE result, where the central region remains strongly peaked.
The qualitative agreement between koopmons and the quantum reference is rather remarkable, given the long time scale and the presence of strongly quantum behavior in the Wigner dynamics, which would normally invalidate MQC models.

\begin{figure}
	\centering
	\includegraphics[width=\textwidth]{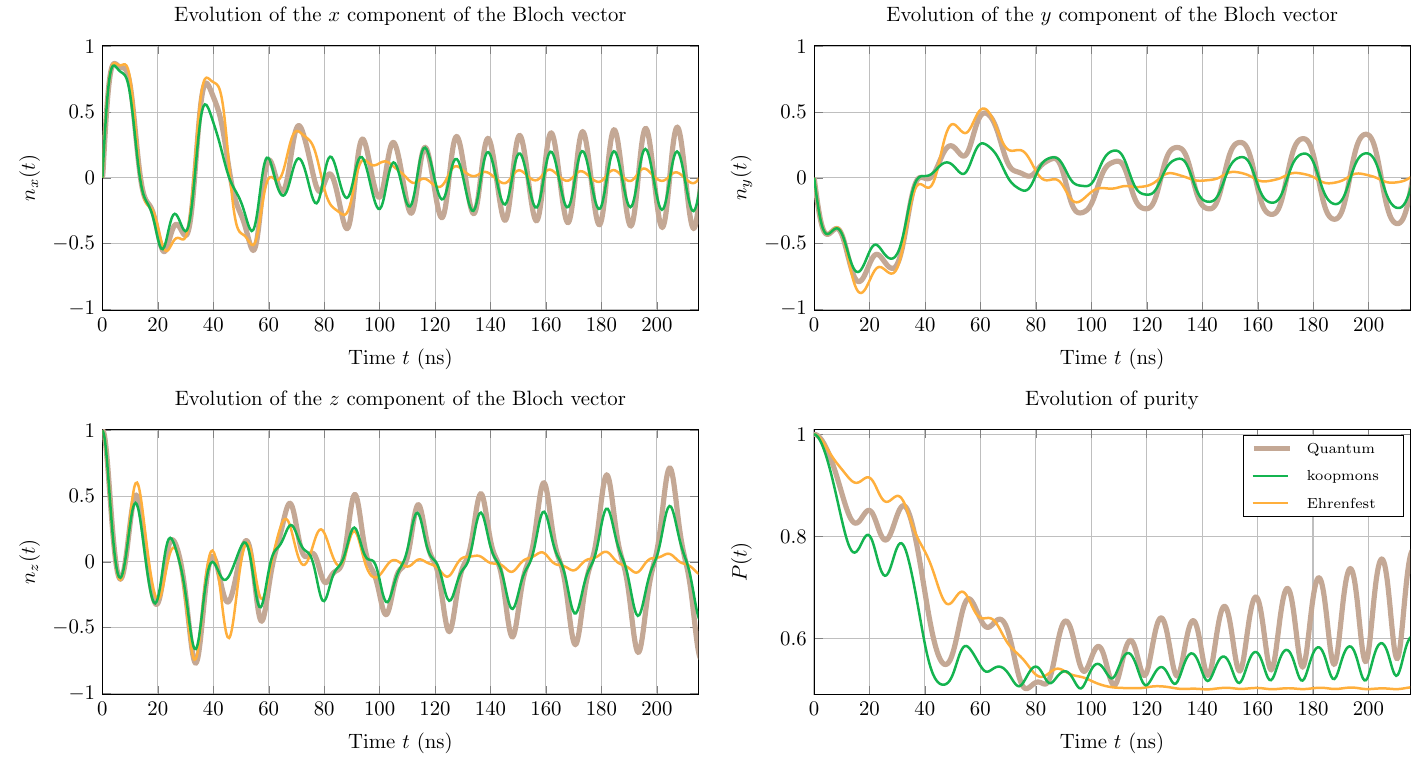}\vspace{-.25cm}
	\caption{\footnotesize
		Time evolution of the Bloch vector components and purity (bottom right) for the non-ballistic, Zeeman-dominated case (GaAs).
	}\label{10-NBZD-B}
\end{figure}
\paragraph{Quantum sector and spin-orbit correlations.}
Figure~\ref{10-NBZD-B} shows the evolution of the Bloch vector and the purity.
All three Bloch components are activated in this test case.
For times up to about $70~\text{ns}$, the MTE curves align well with the quantum reference: the extrema occur at the correct times and the amplitudes are reasonably close.
The koopmons show a similar degree of agreement, even though the mismatch in amplitude is slightly larger.

For longer times ($t\ge 70~\text{ns}$), the koopmons begin to outperform the Ehrenfest method.
The koopmon oscillations follow both the correct phase and the correct amplitude envelope much more accurately than MTE, whose oscillations drift toward zero and eventually settle near a constant value.
A similar picture is visible in the purity plot.
Although the koopmons show somewhat stronger decoherence initially, their oscillation period matches the quantum curve very well over the entire simulation.
In contrast, the MTE purity exhibits a visible drift toward a final purity level close to $1/2$, which does not appear in the quantum result.

\begin{figure}
	\centering
	\includegraphics[width=\textwidth]{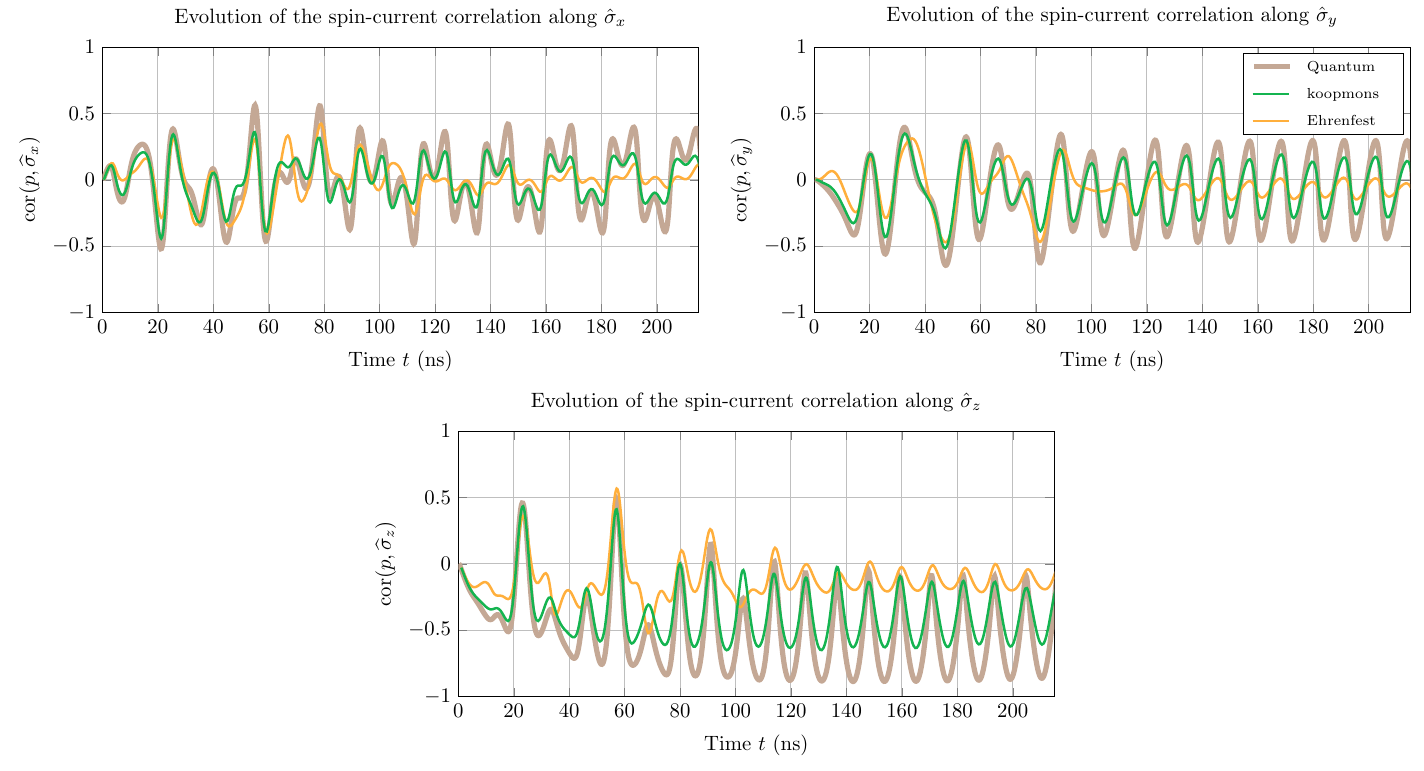}\vspace{-.25cm}
	\caption{\footnotesize
		Time evolution of the spin-momentum correlation components along $\widehat\sigma_x$ (top left), $\widehat\sigma_y$ (top right), and $\widehat\sigma_z$ (bottom) for the non-ballistic, Zeeman-dominated case (GaAs).
	}\label{11-NBZD-SC}
\end{figure}
The evolution of the spin-momentum correlations is shown in Figure~\ref{11-NBZD-SC}.
For the first two components (top left and top right), we see that the koopmon simulation shows very good agreement with the periods of the quantum solution, while the Ehrenfest dynamics only reproduce the oscillatory behavior for times $t\le 70\,\text{ns}$.
At later times, however, especially in the first component, a clear mismatch in the Ehrenfest oscillations becomes visible.
Even though the amplitudes are not captured correctly at larger times, we note that the average value of the oscillations (zero in the first component) is still reproduced reasonably well by the Ehrenfest method in the first two components.
However, this is not the case for the third component (bottom).
The Ehrenfest simulation follows the correct oscillation frequency, but the amplitudes are in complete disagreement with the quantum result and even the mean value is no longer captured.
The koopmons, in contrast, show better accuracy for this component -- both in capturing the period and the amplitude of the oscillations.

\section{Conclusions and perspectives}
In an attempt to probe the applicability of mixed quantum-classical (MQC) models beyond consolidated benchmark problems, we have considered the Ehrenfest and the Koopman models in \eqref{Ehrenfest} and \eqref{HybEq1}-\eqref{HybEq3}, respectively, which succeed in satisfying a series of stringent consistency criteria generally challenging semiclassical Boltzmann-type treatments \cite{Eilenberger,Kailasvuori,Halperin,Gorini}.
In particular, we have implemented the particle closure schemes in \eqref{MFeqs} and \eqref{MFeqs2}-\eqref{koopint}, respectively, to assess the extent to which MQC dynamics succeeds in reproducing the behavior of quantum spin-orbit interaction.
For this purpose, we applied both implementations to compare MQC and fully quantum dynamics of Rashba nanowires in different regimes with realistic semiconductor parameters.
Not only did we compare the dynamics of the orbital and spin degrees of freedom, but we also provided an account of spin-orbit correlations between spin and orbital momentum.
As a general conclusion, this first Rashba validation study shows that the koopmon implementation of the Koopman MQC model in \eqref{HybEq1}-\eqref{HybEq3} captures important coupled spin-orbit features that are missed by the MTE scheme associated to the Ehrenfest MQC model \eqref{Ehrenfest}.
In particular, while the MTE scheme may provide excellent accuracy in the quantum spin dynamics for the ballistic case, the orbital dynamics associated to the Ehrenfest model fail to capture essential quantum features such as wavepacket spreading and splitting in phase space.
For long times, the appearance of strong quantum orbital correlations triggers interference patterns in the Wigner phase space thereby stretching the applicability of MQC approaches.
Nevertheless, we have observed that the koopmon scheme continues to reproduce qualitative long-time features that are otherwise impossible to capture with Ehrenfest dynamics.

It is interesting to observe the overall differences between MQC and quantum dynamics already in the ballistic regime, which is defined by the absence of an external potential.
In the Zeeman-dominated case, the quantum wavepacket splitting in the orbital phase-space is qualitatively captured by the koopmon scheme.
However, in this case the splitting is somewhat blurred: the koopmon density shows low-probability regions between the populated branches, while remaining a nonnegative classical density.
Likewise, in the Rashba-dominated case, regions that are negative in the Wigner representation correspond to particle-depleted regions in the koopmon density.
Thus the correspondence is geometric and qualitative: the koopmon particles tend to occupy regions close to the positive Wigner support while avoiding some regions associated with strong quantum interference.
The koopmon scheme also performs well in capturing the important features of spin dynamics and spin-momentum correlations, although it displays a slight loss of accuracy with respect to the MTE scheme.
The latter remains unable to reproduce the quantum orbital dynamics.

In the non-ballistic regime, the orbital phase-space dynamics leads to the same overall conclusions already obtained in the ballistic case.
This time, however, it is the spin and spin-momentum dynamics that display important differences.
Indeed, in the non-ballistic regime the koopmon method captures several oscillatory features, with substantially higher accuracy than the Ehrenfest model.
This result is far from obvious because non-ballistic dynamics triggers quantum orbital correlations very early on, leading to negative Wigner regions that signal the edge of the applicability range of MQC approaches.
It is remarkable that the koopmon method succeeds in capturing essential dynamical features even in the presence of cat-like states, which are a typical example of non-classical states.

In this work, we focused on momentum-dependent spin-orbit coupling and validated our results against fully quantum SOFT simulations.
Inhomogeneous semiconductor samples with a space-dependent Rashba parameter would involve the simultaneous coupling of both position and momentum, creating non-commuting operators that complicate Fourier-based schemes and the construction of fully quantum benchmarks.
While such systems are a promising possible application of MQC approaches, their numerical treatment and performance assessment are beyond the scope of the present study.
Exploring the applicability of koopmon in these more complex, computationally demanding regimes represents a natural direction for future research.

Moreover, in higher dimensions the koopmon scheme \cite{Bauer24} allows one to reduce the integral in \eqref{koopint} to a combination of pairwise products of two-dimensional integrals, thereby making higher-dimensional systems an attractive direction for future investigation.
Nevertheless, the global evaluation of the integral force terms continues to represent a natural scaling bottleneck.
Possible routes toward improved scalability include random batch strategies for approximating the interaction sums, Hamiltonian Monte Carlo ideas for more efficient phase-space sampling, and parallel evaluation of the backreaction terms.
These acceleration strategies provide concrete directions for future algorithmic development.

Finally, we note that Hamiltonians of the form \eqref{BallHam} are not unique to semiconductor nanowires; for example, they also appear in spin-orbit-coupled Bose--Einstein condensates \cite{LiJiSp11,ZhMoBuEnZh16}.
In these systems, nonlinearities arise that depend solely on the position-space probability density while leaving the spin-orbit interaction operator unaffected.
Adapting the koopmon scheme to navigate these nonlinear landscapes represents a compelling frontier for future research.

\paragraph{Acknowledgments.}
We thank Denys Bondar, Ignacio Franco, and Eran Ginossar for several insightful discussions and suggestions during the development of this work.
The work of PB in this project was funded by the National Science Centre, Poland, (NCN) project no. 2019/34/E/ST1/00390.
GM acknowledges financial support by the Interdisciplinary Thematic Institute QMat, as part of the ITI 2021-2028 program of the University of Strasbourg, CNRS and INSERM, IdEx Unistra (ANR 10 IDEX 0002), SFRI STRAT'US project (ANR 20 SFRI 0012), and ANR-17-EURE-0024 under the framework of the French Investments for the Future Program.
CT acknowledges financial support by the Leverhulme Trust Research Project Grant RPG-2023-078.

%

%

\end{document}